\documentclass{jfm}

\usepackage{amssymb}
\usepackage{amsmath}
\usepackage{color}
\usepackage{natbib}

\usepackage{mathrsfs}
\usepackage{esint}
\usepackage{graphicx}
\usepackage{epstopdf, epsfig}
\usepackage{color}

\newcommand{\be}{\begin{equation}}
\newcommand{\ee}{\end{equation}}
\def\ve{{\varepsilon}}
\def\cU{{\cal U}}
\newcommand{\half}{\mbox{$\frac{1}{2}$}}

\def\cN{{\cal N}}

\shorttitle{Nonlinear dynamics of forced baroclinic critical layers}
\shortauthor{C. Wang and N. J. Balmforth}

\title{Nonlinear dynamics of forced baroclinic critical layers}
\author{Chen Wang\aff{1}
  \corresp{\email{chenwang@math.ubc.ca}}
  \and Neil J. Balmforth\aff{1}
  }

\affiliation{\aff{1}Department of Mathematics, University of British Columbia,
Vancouver, BC V6T 1Z2, Canada}

\begin{document}

\maketitle

\begin{abstract}
\noindent
      In this paper, we study the forcing of baroclinic critical levels,
      which arise in stratified fluids with horizontal shear flow along the
      surfaces where the phase speed of a wave relative to the mean flow
      matches a natural internal wavespeed.
      Linear theory predicts the baroclinic critical layer dynamics
      is similar to that of a classical critical layer, characterized by the
      secular growth of flow perturbations over a region of decreasing width.
      By using matched asymptotic expansions, we construct a nonlinear
      baroclinic critical layer theory to study how the flow
      perturbation evolves once they enter the nonlinear regime. A key
      feature of the theory is that, because the location of the
      baroclinic critical layer is determined by the streamwise wavenumber,
      the nonlinear dynamics filters out harmonics and the modification to the
      mean flow controls the evolution. At late times, we show that
      the vorticity begins to focus into yet smaller
      regions whose width decreases exponentially with time,
       and that the addition of dissipative effects
        can arrest this focussing to create a drifting coherent structure.
      Jet-like defects in the mean horizontal velocity
      are the main outcome of the critical-layer dynamics.
\end{abstract}

%\begin{keywords}
%Authors should not enter keywords on the manuscript, as these must be chosen by the author during the online submission process and will then be added during the typesetting process (see http://journals.cambridge.org/data/\linebreak[3]relatedlink/jfm-\linebreak[3]keywords.pdf for the full list)
%\end{keywords}

\section{Introduction}
          A centrepiece in the theory of inviscid shear flow is the classical
          critical level, where the phase speed $c$
          of a steady wave matches the local mean flow speed $U$.
          In linear theory, the levels where $c=U$
          become singular, demanding the inclusion of the weak
          effects of unsteadyness, nonlinearity or viscosity
          \citep{Maslowe86}. Although these inclusions can remove
          the singularity of the linear inviscid theory, perturbations
          to the flow can still develop strongly in the neighbourhood
          of the critical levels, creating distinctive flow structures
          and rearrangements within the so-called critical layers
          that may subsequently break down to generate mixing and
          turbulence. In this vein, \citet{Stewartson78} and
          \citet{Warn76, Warn78} studied the nonlinear dynamics
          of the critical layers of forced
          Rossby waves. They found that steady waves developed over the
          bulk of the shear flow, but that the critical layer remained
          unsteady, exciting mean-flow corrections
          and all the harmonics of the original
          wavenumber, and twisting up the background vorticity
          into Kelvin cat's eye pattern.
          A similar scenario exists for the critical layers of internal
          gravity waves travelling vertically through stratified shear flow,
          with important repercussions on wave breaking,
          momentum transport and mixing
          in the atmosphere \citep{Booker67,Brown80,Brown82a,Brown82b}.

          If the flow is stratified vertically but sheared horizontally, then
          a new type of critical level appears in the linear inviscid wave
          theory. The new critical levels arise along the surfaces where the
          phase speed relative to the background shear flow matches a
          characteristic velocity of gravity waves; \textit{i.e.}
          $c-U=\pm N/k$, where $N$ is the buoyancy frequency and $k$ is the
          streamwise wave number. Existing literature on these
          `baroclinic critical levels', has
          mainly focused on the propagation of linear wave packets.
          Using ray-tracing theory, \citet{Olbers81}, \citet{Basovich84} and
          \citet{Badulin85} found that wave packets slow down as they approach
          the baroclinic critical level, never reaching it.
          Simultaneously, the wave amplitude and cross-stream wavenumber grow
          indefinitely, indicating that linear theory eventually fails
          in a wave-trapping process like that found earlier for
          classical critical levels \citep{Bretherton66}. \citet{Staquet02}
          and \citet{Edwards05} performed numerical simulations to study the
          nonlinear evolution during trapping, concluding that
          the trapped waves may either break into small-scale
          turbulence or be dissipated by dispersion, viscosity and diffusion.
          More related to the current work
            is the study by \cite{boulanger}, who explored the
            analogues of baroclinic critical levels in stratified,
            titled vortices, and resolved the resulting singularities
            by introducing viscosity.

        Baroclinic critical layers have also featured heavily
          in recently reported computations of
          three dimensional rotating stratified shear flows with
          self-replicating vortices \citep{Marcus15, Marcus16, Marcus18}.
          The replication process involves the forcing of baroclinic
          critical layers by internal waves excited by an initial vortex;
          large-amplitude re-arrangments forced in these layers then roll
          up to create new votices, which in turn shed more internal
          waves to repeat a cycle. The self-replication eventually
          filled the computational domain with localized vortical structures,
          which was suggested to be trigger for the angular momentum transport
          required to drive accretion in astrophysical disks that are
          too cool to suffer the magneto-rotational instability.

          The aim of the present paper is to theoretically study the
          evolution of forced baroclinic critical layers, following the
          paradigm of \citet{Stewartson78} and \citet{Warn76, Warn78}
          for Rossby waves, or \citet{Booker67} and
          \citet{Brown80,Brown82a,Brown82b} for internal waves in
          stratified shear flow. The linear dynamics of a forced
          baroclinic critical layer is expected to be similar to that
          of a classical critical layer, owing to the similarity of
          the singularities in the
          linear wave equations. However, the subsequent nonlinear evolution
          is likely to be very different because the location of
          the baroclinic critical
          level itself is dictated by the streamwise wave number, which is
          different among all the harmonics of the original wave. This
          suggests that they cannot feature in the nonlinear dynamics
          within the baroclinic critical layer, unlike in
          classical critical layer theory.

          The layout of the paper is as follows: in \S 2, we give the model
          and governing equations of the problem. In \S 3, we solve the linear
          problem explicitly and draw out structure that first develops
          within the baroclinic critical level. In \S 4, we
          extend the analysis by considering weakly nonlinear perturbations,
          which allows us to determine the time and length scales that
          characterize the
          nonlinear critical layer. This leads us, in \S 5,
          to derive a reduced model of nonlinear dynamics via a matched
      asymptotic expansion. We then present numerical solutions of the reduced
      model and a further asymptotic analysis of them.
       We explore the effects of dissipation in the baroclinic
        critical layer in \S 6, and then discuss the implications
        of the results and the relation to previous and
          future work in \S 7.

\section{Model and governing equations}

%As illustrated in figure \ref{F 1}, w
We consider
forced disturbances to an unbounded horizontal shear flow, orientated
in the $x-$direction with a constant shear rate $\Lambda>0$
in the $y-$direction.
The domain rotates around the vertical axis at angular velocity $\Omega$, and
the fluid is stratified in $z$ with constant buoyancy frequency $N$. Waves are driven into the shear flow by a wavemaker that
  we locate along $y=0$. This forcing has
%The disturbance is forced by switching on a weak periodic forcing
%at $y=0$, which generates a jump in the tangential horizontal velocity
%but not in the normal velocity. The forcing has
the streamwise and vertical wavenumbers, $k_x$ and $k_z$, respectively. The baroclinic critical levels are located at $y=\pm N/(\Lambda k_x)$.  The sketch of the model is shown in figure 1.

\begin{figure}
 \begin{center}
   \includegraphics[width=0.8\linewidth]{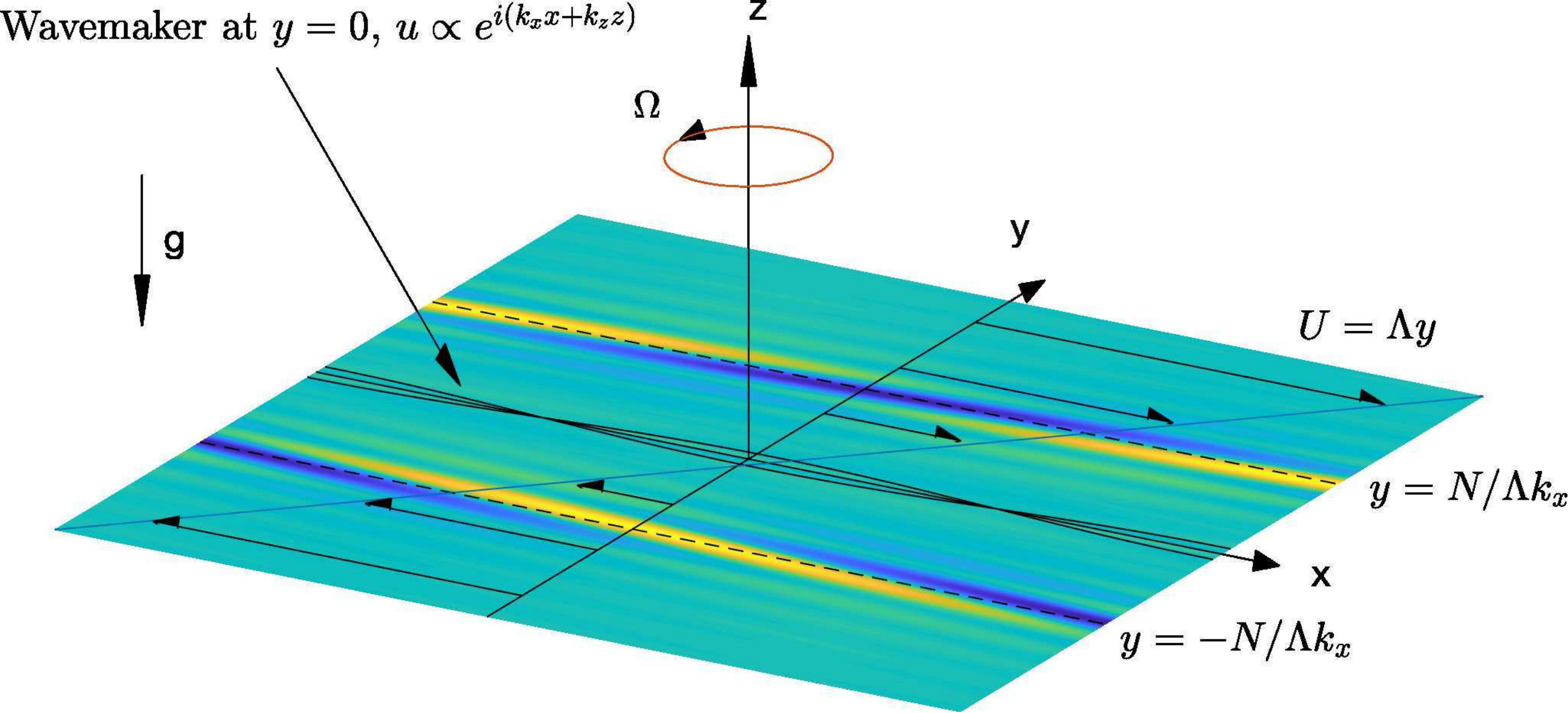}
   \caption{Sketch of the model. A wavemaker with wavenumber $k_x$ and $k_z$ is imposed at $y=0$, and baroclinic critical levels are forced at $y=\pm N/(\Lambda k_x)$, corresponding to dimensionless locations $\pm\mathcal{N}$, where $\cN=N\Lambda^{-1}$. The shading represents a
       rendering of the density perturbation based on the linear theory of
   \S 3.}
   \label{F 1}
 \end{center}
\end{figure}

{\color{blue}
%As shown in figure 1, such a forcing approximates a thin periodic vortex
%sheet, unlike in the studies of \citet{Stewartson78} and \citet{Booker67},
%where a wavy boundary forced the normal velocity. Here we choose the vortex
%sheet to render the problem closer to the that considered by \citet{Marcus13}.
%A boundary at $y=0$ could also make the basic flow in figure \ref{F 1}
%unstable \citep{Wang18} and thus make the problem more complicated.
%Aside from this, however, the precise form of
%forcing of the wave should not affect
%the qualitative behavior of the baroclinic critical layers.
}
We work with a dimensionless version of the governing fluid
equations in which length, time, velocity, pressure and density perturbations
are scaled by $k_x^{-1}$, $\Lambda^{-1}$, $\Lambda k_x^{-1}$,
$\rho_0\Lambda^2 k_x^{-2}$ and $\rho_0\Lambda^2/(k_xg)$, respectively.
Here, $\rho_0$ is a reference density and $g$ is gravity.
We employ the Boussinesq approximation and, for the most part of our study,
neglect viscosity and diffusion in view of the large spatial scales
that characterize geophysical and astrophysical flows.
At the end of the
work, we briefly explore the effect of diffusion.
The perturbations to the
  velocity $(u,v,w)$, pressure $p$ and perturbation density $\rho$
then satisfy
\begin{align}
u_t+yu_x+(1-f)v+uu_x+vu_y+wu_z&=-p_x, \label{1.1}\\
v_t+yv_x+fu+uv_x+vv_y+wv_z&=-p_y, \label{1.2}\\
w_t+yw_x+uw_x+vw_y+ww_z&=-p_z-\rho,  \label{1.3}\\
\rho_t+y\rho_x-\cN^2 w+u\rho_x+v\rho_y+w\rho_z&=0 \label{1.4},\\
u_x+v_y+w_z&=0, \label{1.5}
\end{align}
where subscripts represent partial derivatives
and we have introduced the dimensionless Coriolis parameter
$f=2\Omega/\Lambda$ and buoyancy frequency $\cN=N\Lambda^{-1}$.
Because our interest lies in the forcing of the baroclinic critical layers of
an internal wave, we consider basic flows that are linearly
stable to prevent unstable modes from dominating the dynamics.
Centrifugal instabilities arise when $0<f<1$
\citep{Emanuel94}, so we set $f>1$ or $f<0$ to eliminate them;
strato-rotational instability is not present because it requires
reflective boundaries \citep{Yavneh01, Wang18} which are absent here.

Initially, there is no disturbance, implying
$u=v=w=\rho=p=0$ at $t=0$. The
 wavemaker is then switched on
  to excite waves with baroclinic critical levels.
  To idealize the forcing and formulate a concise mathematical problem,
  we assume that the wavemaker introduces a time-independent
  jump in the tangential horizontal velocity at $y=0$,
  but not in the normal velocity. That is, we impose the jump
  conditions,
  \begin{equation}
    %[u]_{y=0-}^{y=0+}
   u|_{y=0+}- u|_{y=0-} =\varepsilon_0\exp(\mathrm{i}x+\mathrm{i}mz)+\mathrm{c.c.},\quad
   v|_{y=0+}=v|_{y=0-}, %\quad t\ge 0,
  \label{2.6}
\end{equation}
where $\varepsilon_0$ represents the strength,
$m=k_z/k_x$, c.c. represents the complex conjugate,
and the $\pm$ superscripts indicate the limits from either side.
This forcing approximates a thin, spatially periodic vortex
sheet. %\textcolor[rgb]{0.00,0.59,0.00}{ which can be viewed as an idealization of the periodic Gaussian vortices considered by \citet{Marcus13}}. %rendering the problem similar to that considered by \citet{Marcus13},
%in which a vortical structure acts as the wavemaker.
In the numerical simulation of \citet{Marcus13}, waves were forced by a periodic array of localized Gaussian vortices. Our forcing therefore represents an idealization of their model in that we consider the leading-order Fourier component while neglecting the evolution and cross-stream thickness of the
forcing.
The configuration
is slightly different to that in
the studies of \citet{Stewartson78} and \citet{Booker67},
where a wavy boundary forced the normal velocity. The current configuration
implies that waves are generated at $y=0$ and develop with baroclinic
critical levels to either side (although simplifications are afforded
by the symmetry described presently). Had we placed the wavemaker
along a boundary at $y=0$, only one critical level would have
featured, but the wall may also make the basic flow linearly
unstable \citep{Wang18}. Other idealizations include wavemakers
that gradually switch on \citep{Beland76}, that
generates disturbances with finite phase speed (displacing the
baroclinic critical levels), or that with finite
thickness (as for the vortices of Marcus {\it et al.}).
Nevertheless, the precise form of
forcing of the wave is not expected to affect
the qualitative dynamics of the baroclinic critical layers,
a feature on which we elaborate further later.

Note that
the system in (\ref{1.1})-(\ref{2.6}) is invariant under the transformation,
\begin{equation}
(u,v,w,\rho) \to -(u,v,w,\rho) \quad {\rm and} \quad p \to p,
\quad {\rm for} \quad (x,y,z) \to -(x,y,z).
\label{1.10a}
\end{equation}
This observation permits us to solve the problem only in $y>0$,
and therefore consider only one baroclinic crtical layer, then
generate the solution in $y<0$ using the implied symmetry conditions.

Also,
combining (\ref{1.1})-(\ref{1.5}), we may derive an equation for the vertical
component of vorticity:
\begin{equation}
\frac{D}{Dt}(v_x-u_y)-\cN^{-2}(f-1+v_x-u_y)\frac{\partial}{\partial z}\frac{D\rho}{Dt}+w_xv_z-w_yu_z=0, \label{1.6}
\end{equation}
where
\begin{equation}
\frac{D}{Dt}=\frac{\partial}{\partial t}+(y+u)\frac{\partial}{\partial x}+v\frac{\partial}{\partial y}+w\frac{\partial}{\partial z}.
\end{equation}

\section{Linear theory}
The linearized governing equations are
\begin{align}
u_t+yu_x+(1-f)v&=-p_x,
\label{2.1}\\
v_t+yv_x+f u&=-p_y,
\label{2.2}\\
w_t+yw_x+\rho&=-p_z,
\label{2.3}\\
\rho_t+y\rho_x-\cN^2 w&=0,
\label{2.4}\\
u_x+v_y+w_z&=0.
\label{2.5}
\end{align}
The linearized equation of (\ref{1.6}) reduces to a conservation law of
potential vorticity, $q_t+yq_x=0$, or, given that $q=0$ everywhere at $t=0$,
\begin{equation}
  q=(f-1)\rho_z-\cN^2(v_x-u_y)=0 .
\label{2.7}
\end{equation}

%\begin{equation}
%q=(f-1)\rho_z-\cN^2(v_x-u_y)=0, \quad y\neq 0,\quad t\ge 0.  \label{2.7a}
%\end{equation}

In the absence of linear instability,
the forcing (\ref{2.6}) drives a steady wave response
throughout the bulk of the flow
(as can be established by solving the initial-value problem using
Laplace transforms, and then performing a large-time asymptotic analysis,
following \citet{Warn76} and \citet{Booker67}).
Near the baroclinic critical levels, however, the flow remains unsteady,
requiring a finer analysis of those regions similar to
that used by \citet{Stewartson78}.

\subsection{The steady wave response outside the baroclinic critical layers}

The steady wave solution outside the critical layers takes the form:
\begin{equation}
  \left(u,v,w,p,\rho \right)=\left[\hat{u}(y),\hat{v}(y),\hat{w}(y),
    \hat{p}(y),\hat{\rho}(y)
    \right]\exp\left(\mathrm{i}x+\mathrm{i}mz\right)+\mathrm{c.c.}
    \label{2.8}
\end{equation}
Substituting (\ref{2.8}) into (\ref{2.1})-(\ref{2.5}), one can derive an
equation for $\hat{p}(y)$,
\begin{equation}
\hat{p}''-\frac{2y}{y^2-f(f-1)}\hat{p}'-\left[\frac{y^2-f(f+1)}{y^2-f(f-1)}+m^2\frac{y^2-f(f-1)}{y^2-\cN^2}\right]\hat{p}=0, \label{2.9}
\end{equation}
with
\refstepcounter{equation}
$$
\hat{u}=\frac{(f-1)\hat{p}'-y\hat{p}}{y^2-f(f-1)},\quad
\hat{v}=\frac{\textrm{i}(y\hat{p}'-f\hat{p})}{y^2-f(f-1)}, \quad
\hat{w}=-\frac{my\hat{p}}{y^2-\cN^2},\quad
\hat{\rho}=\frac{\mathrm{i}m\cN^2\hat{p}}{y^2-\cN^2}.
\eqno{(\theequation{a,b,c,d})}\label{2.11}
$$
({\it cf.} Vanneste \& Yavneh, 2007). Note that the singularities at
$y^2=f(f-1)$ in (\ref{2.9}) and (\ref{2.11}) are removable.
The baroclinic critical levels $y=\pm\cN$, however, are true singular points. The Frobenius solutions near $y=\cN$ are,
\begin{align}
  \hat{p}_{A} &= 1 - \frac{m^2[\cN^2-f(f-1)]}{2\cN}(\cN - y)\log|\cN - y|
  - \alpha (\cN - y)+...  %O((y-y_{B\pm})^2,(y-y_{B\pm})^2\log|y-y_{B\pm}|)
\label{2.13a}
\\
\hat{p}_{B} &=  y - \cN + ...
\label{2.13b}
\end{align}
where $\alpha$ is determined by the condition that
$\hat{p}_{A}\rightarrow 0$ as $y\rightarrow  \infty$.
In terms of these Frobenius solutions, we express $\hat{p}$ for $y>0$ by
\refstepcounter{equation}
\begin{equation}
  \hat{p}=\left\{
  \begin{array}{ll} A_{L}\hat{p}_{A}, &
    y> \cN, \\
 A_{L}\hat{p}_{A}+ B_{L}\hat{p}_{B}, & 0< y< \cN,
 \end{array}\right.
 \label{2.144}
\end{equation}
where $A_{L}$ and $B_{L}$ are constants.

Although $\hat{p}$ is bounded for  $y\rightarrow \cN$,
the amplitudes of the velocity,
$(\hat{u},\hat{v},\hat{w})$, and density, $\hat{\rho}$, all diverge,
signifying that the steady wave solution fails at the critical levels.
In particular, we observe that
\begin{equation}
  \hat{p}\to A_{L}, \qquad
  \hat\rho \to \frac{\mathrm{i}m\cN A_{L}}{2(y-\cN)}
  \label{match1}
\end{equation}
and
\begin{equation}
  \hat{u}\to  \left[\frac{m^2(f-1)}{2\cN}(\log|\cN-y| + 1)
  + \frac{\alpha(f-1)-\cN}{\cN^2-f(f-1)}\right]A_{L}
  + \left\{ \begin{array}{ll}
    0 & y>\cN, \cr
    \frac{f-1}{\cN^2-f(f-1)}B_{L} & y<\cN,
    \end{array}\right.
  \label{match2}
\end{equation}
for $y\to\cN$.

\subsection{The linear critical layers}

We now focus on the
baroclinic critical layer at $y=\cN$.
Here, we search for an unsteady solution
depending on the long timescale $T=\delta t$ and with the short spatial scale
$Y=(y-\cN)/\delta$, where $\delta\ll1$ is a small parameter
organizing an asymptotic expansion.
We then set
\begin{equation}
  \left(u,v,w,p,\rho \right)=\left[\widetilde{u}(Y,T),\widetilde{v}(Y,T),
    \delta^{-1} \widetilde{w}(Y,T),
    A_{L},\delta^{-1} \widetilde{\rho}(Y,T)
    \right]\exp\left(\mathrm{i}x+\mathrm{i}mz\right)+\mathrm{c.c.},
 \label{2.14}
\end{equation}
in view of the limits in (\ref{match1})-(\ref{match2}).

Combining (\ref{2.3}) and (\ref{2.4}) to eliminate $w$, then substituting
in (\ref{2.14}) now gives, to leading order in $\delta$,
\begin{equation}
  \left(\frac{\partial}{\partial T}+\mathrm{i}Y\right)\widetilde{\rho} =
  - \frac{1}{2}m\cN A_{L}.
%  \left[\left(\frac{\partial}{\partial t}+\textrm{i}y\right)^2+R\right]\widetilde{\rho}=-\textrm{i}Rm\widetilde{p}.
  \label{2.15}
\end{equation}
In the early stage of linear evolution, $t\sim O(1)$, $\rho\sim O(1)$,  so we have the initial condition $\widetilde{\rho}\rightarrow 0$ as $T\rightarrow0$, which yields
\begin{equation}
    \widetilde{\rho} = -\frac{1}{2} \textrm{i} m \cN A_{L}
\frac{e^{-\mathrm{i}YT}-1}{Y} , \label{2.18}
\end{equation}
Hence
\begin{equation}
    \rho = -\frac{1}{2} \textrm{i} m \cN A_{L} t
  \left[\frac{e^{-\mathrm{i}(y-\cN)t}-1}{(y-\cN)t}\right]e^{\mathrm{i}x+\mathrm{i}mz}+\mathrm{c.c.}   \label{2.19}
\end{equation}
This solution has a spatial structure dependent on
the self-similar combination $t(y-\cN)$.
Hence, the amplitude grows linearly and the width of the critical layer
shrinks with time.
% Didn't understand the comment about $\delta$ here;
% there is a time-independent choice for it in the nonlinear analysis
% and one cannot rescale as we have done if $\delta$ depends on time.
% An alternative would have been to search for a long-time
% simiarity solution, which I think is what you are getting at;
% but that's a little different from the apprach currently taken.

Next, the main balance in (\ref{2.7}) implies that
$\widetilde{u}_Y \sim -\mathrm{i}m(f-1)\cN^{-2}\widetilde\rho$, or
 \begin{equation}
 \widetilde{u}_Y=-\frac{m^2(f-1)A_{L}}{2\cN}\frac{e^{-\mathrm{i}YT}-1}{Y}.\label{2.21}
 \end{equation}
 But the limits of the steady wave response in (\ref{match2}) imply that
 $\widetilde{u}$ jumps by an amount $(f-1)B_L/[\cN^2-f(f-1)]$
 across the baroclinic critical layer. Hence,
% Integrating, and then demanding that the solution match to the
% $y>\cN$ limit in (\ref{match2}), now gives
% \begin{equation}
%   \widetilde{u}=\frac{m^2(f-1)A_{L}}{2\cN}
%   \left[\log|\delta Y|+\fint^{\infty}_Y e^{-\mathrm{i}Y'T} \frac{\textrm{d}Y'}{Y'}
%     + 1 \right]+\frac{(f-1)\alpha-\cN}{\cN^2-f(f-1)}A_{L} ,
% \label{2.22}
% \end{equation}
%and the match to the limit in (\ref{match2}) for $y<\cN$ furnishes
%a relation for $B_{L}$: %By conducting the same analysis for the
%critical layer at $y=-\cN$, we eventually find
\begin{equation}
  B_{L}=-\frac{m^2A_{L}[f(f-1)-\cN^2]}{2\cN}
  \lim_{L\to\infty}\int_{-L}^{L} (e^{-\mathrm{i}YT}-1)\frac{\textrm{d}Y}{Y}
  = \textrm{i}\pi\frac{m^2[f(f-1)-\cN^2]}{2\cN}A_{L}. \label{2.26}
\end{equation}
({\it cf.} Stewartson 1978).
%Here, the decoration of the integral indicates the principal value.

\subsection{Closure}

\begin{figure}
\begin{center}
   \includegraphics[width=0.9\linewidth]{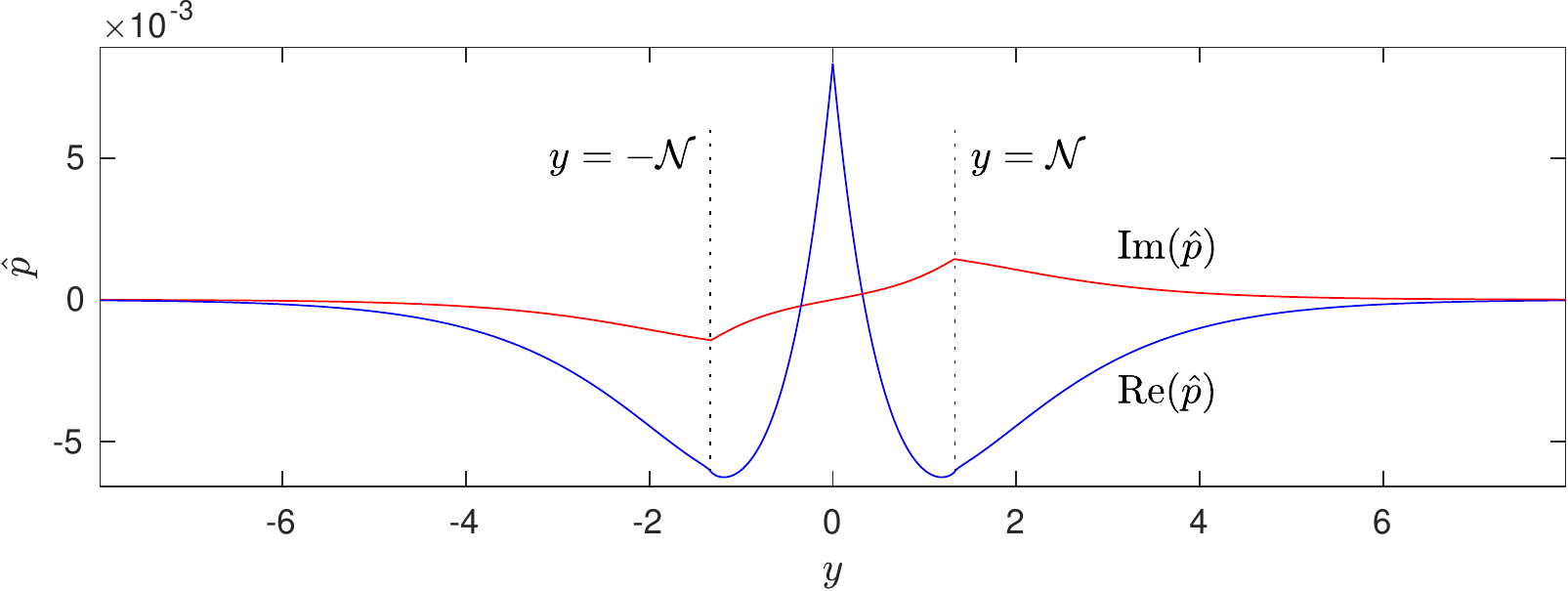}
   \caption{
     Steady-wave solution $\hat{p}$ under a forcing imposed at $y=0$, with $m=0.5$, $\cN=4/3$, $f=4/3$,
     $\varepsilon_0=0.05$ ({\it cf.} Marcus \textit{et al}. 2013). Baroclinic critical levels $y=\pm\cN$ are indicated.
  }
   \label{F2}
 \end{center}
\end{figure}

We can now apply the forcing condition to close the problem.
The symmetry property (\ref{1.10a}) applied to the steady wave (\ref{2.8})
indicates that
\begin{equation}
    [\hat{u}(y),\hat{v}(y),\hat{w}(y),\hat{\rho}(y)]=-[\hat{u}(-y),\hat{v}(-y),\hat{w}(-y),\hat{\rho}(-y)]^*, \quad \hat{p}(y)=\hat{p}(-y)^*, \label{3.22}
\end{equation}
  where the superscript $^*$ represents the complex conjugate.
Hence, substituting the steady wave solution into the jump condition
(\ref{2.6}) representing the forcing, we arrive at
%\begin{align}
\refstepcounter{equation}
  $$
  (A_L-A_L^*)\hat{p}_{A}(0)+(B_{L}-B_{L}^*)\hat{p}_{B}(0) =0,
  $$
  $$
\quad \quad (A_{L}+A_{L}^*)\hat{p}'_{A}(0)+(B_{L}+B_{L}^*)\hat{p}'_{B}(0) =-f\varepsilon_0.
 \eqno{(\theequation{a,b})}
 \label{forc1}
 $$
%\end{align}
Exploiting (\ref{2.26}), we obtain
\begin{equation}
  A_{L}=-
  \left. \frac{f\varepsilon_0(\hat{p}_{A}-\textrm{i}\beta\hat{p}_{B})}
       {2(\hat{p}_{A}\hat{p}_{A}'+\beta^2\hat{p}_{B}\hat{p}_{B}')}\right|_{y=0},
  \quad \beta=\frac{\pi m^2[f(f-1)-\cN^2]}{2\cN}.  \label{2.29}
\end{equation}
The amplitude of the pressure perturbation at the
critical layer is therefore
\begin{equation}
  \left.\varepsilon=|A_{L}|=
  \frac{|f\varepsilon_0|\sqrt{\hat{p}_A^2+\beta^2\hat{p}_B^2}}
     {2\left|\hat{p}_A\hat{p}_A'+\beta^2\hat{p}_B\hat{p}_B'\right|}\right|_{y=0}.
       \label{2.30}
\end{equation}
A sample steady wave solution is plotted in figure {\ref{F2}}.

  Note that equations
  (\ref{forc1})-(\ref{2.30}) appear to become trivial if $f=0$,
  suggesting that rotation
  is essential to the forcing of the baroclinic critical layer. In fact,
  a deeper analysis of the Frobenius solutions demonstrates that
  this is not the case, because
  $\hat{p}_A'(0)$ and $\hat{p}_B'(0)$ become $O(f)$ in this limit,
  and the closure relation in (\ref{forc1}) remains non-trivial.
  Consequently, in the model, we may take the limit $f\to0$, highlighting
  how rotation is not an essential ingredient to the dynamics.

The same feature does not apply to the vertical wavenumber or stratification,
which control the secular growth inside the
critical layer, as seen in (\ref{2.18}) and (\ref{2.21});
without either a vertical dependence in the forcing or
stratification, there is no baroclinic critical-layer dynamics.
Note that, despite appearances,
the limit  $\cN\rightarrow0$ in  (\ref{2.21}) is not problematic:
further analysis of $\hat{p}_A$ and $\hat{p}_B$
indicates that $|A_L|\sim\mathcal{N}/\log\cN$ for $\cN\to0$,
and so the secular growth in the critical layer is eliminated
in this limit.

It is also noteworthy that,
in the limit that any of the parameters $m$, $f$, or $\cN$
are large, the disturbance decays exponentially
from the forcing to the baroclinic
critical levels ({\it cf.} \citet{Vanneste07} and \citet{Wang18}).
The amplitude ratio $\varepsilon/\varepsilon_0$ then becomes
exponentially small, and
the secular growth in the critical layer is much weakened.

\section{The weakly nonlinear critical layer}
We now advance beyond linear theory and perform a weakly nonlinear
expansion by setting
\begin{eqnarray}
  (u,v,w,\rho,p)&=&
  \varepsilon \left\{
     [u_1(Y,T),v_1(Y,T),\delta^{-1}w_1(Y,T),\delta^{-1}\rho_1(Y,T),p_1(Y,T)]
  e^{\mathrm{i}x+\mathrm{i}mz}+\mathrm{c.c.}\right\}
  \nonumber  \\
  &&+\varepsilon^2[u_0(Y,T),v_0(Y,T),w_0(Y,T),\rho_0(Y,T),p_0(Y,T)]
 \label{3.1}
 \\
  &&+\varepsilon^2\left\{ [u_2(Y,T),v_2(Y,T),w_2(Y,T),\rho_2(Y,T),p_2(Y,T)]
 e^{2(\mathrm{i}x+\mathrm{i}mz)}+\mathrm{c.c.}\right\},
   \nonumber
\end{eqnarray}
focussing upon the critical layer with $y=\cN+\delta Y$.
The scaling of the fundamental Fourier component follows the linear
critical layer theory outlined above, and we have
$\varepsilon [u_1,v_1,w_1,\rho_1,p_1]\to
[\widetilde{u},\widetilde{v},\widetilde{w},\widetilde{\rho},A_L]$
at early times ($T\ll1$). The goal of the current
section is to identify the timescale and width
of the critical layer (as dictated by the small parameter $\delta$)
for which the mean flow correction
and first harmonic reach sufficient strength to
modify the evolution of fundamental mode. This connects $\delta$
to the amplitude parameter $\varepsilon$, establishing
the scalings of the nonlinear critical layer.

\subsection{Mean-flow response}

The mean-flow component of (\ref{1.5}) gives $v_{0Y}=0$, which implies
$v_{0}=0$ since the mean flow response decays outside the critical layer.
The streamwise mean-flow velocity $u_0$ is described by the $j=0$ component
of (\ref{1.1}), which is
\begin{equation}
  \frac{\partial u_{0}}{\partial T} = \delta^{-2}(\textrm{i}mw_1u_1^*-v_1^*u_{1Y})
  +\mathrm{c.c.}
   \label{3.4aa}
\end{equation}
%\textcolor[rgb]{0.00,0.07,1.00}{ With $u_1$, $v_1$ and $w_1$ satisfying conservation equation (\ref{1.5}) and with $q=0$ for linear disturbances, the evolution law of pseudomomentum (\ref{1.11}) is equivalent to (\ref{3.4aa}) so $u_0$ can be described by the pseudomomentum:}
%\begin{equation}
%u_0=\textsf{p}=\frac{1}{R}(\textrm{i}w_1\rho_1^*-\textrm{i}mu_1\rho_1^*)+\mathrm{c.c.}\sim O(\varepsilon^2t^2).  \label{3.4a}
%\end{equation}
To leading order in $\delta$,
the mean-flow components of (\ref{1.3}) and (\ref{1.4}) are,
\begin{equation}
    \rho_0 = - \delta^{-2} v_1^*w_{1Y}+\mathrm{c.c.}, \label{3.4}
\end{equation}
\begin{equation}
  - \cN^2 w_0 = - \delta^{-2}(v_1^*\rho_{1Y}+\textrm{i}mw_1^*\rho_1) +
  \mathrm{c.c.} \label{3.5}
\end{equation}
Thus, $u_0$, $w_0$, and $\rho_0$ are all $O(\delta^{-2})$.

\subsection{First harmonic}
The largest first harmonic
components of (\ref{1.3}), (\ref{1.4}), (\ref{1.5}) and (\ref{1.6}) indicate
that
\begin{align}
  2\mathrm{i}\cN w_2+\rho_2+2\mathrm{i}mp_2
  &= - \delta^{-2} (v_1w_{1Y}+\textrm{i}mw_1^2), \label{3.9}\\
  2\mathrm{i}\cN\rho_2-\cN^2 w_2 &= -\delta^{-2}
  (v_1\rho_{1Y}-\textrm{i}mw_1\rho_1).  \label{3.10}\\
\delta^{-1} v_{2Y} + 2\mathrm{i}mw_2 &= 0.  \label{3.14}\\
  \frac{u_{2Y}}{\delta}
  + \frac{2\mathrm{i}m(f-1)}{\cN^2}\rho_2
  &=  \frac{\mathrm{i}\delta^{-2}}{2\cN}\left[
    \frac{2\mathrm{i}m(f-1)}{\cN^2}(v_1\rho_{1Y}+\textrm{i}mw_1\rho_1)
  + (v_1u_{1Y}+\textrm{i}mw_1u_1)_Y \right]
  .
  \label{3.13}
\end{align}
However, (\ref{1.2}) demands that $p_2=O(\delta u_2,\delta v_2)$
and so $p_2$ is much smaller than $w_2$ or $\rho_2$.
Hence,
\begin{align}
  w_2 &= \frac{\delta^{-2}}{3\cN^2}
  [2\mathrm{i}\cN(v_1w_{1Y}+\mathrm{i}mw_1^2)-v_1\rho_{1Y}-
    \textrm{i}mw_1\rho_1], \label{3.11} \\
  \rho_2 &=\frac{\delta^{-2}}{3\cN}[\cN(v_1w_{1Y}+\textrm{i}mw_1^2)+2\mathrm{i}
    (v_1\rho_{1Y}+\textrm{i}mw_1\rho_1)], \label{3.12}
\end{align}
which are $O(\delta^{-2})$, whereas
$u_2$ and $v_2$ are $O(\delta^{-1})$.

\subsection{Weakly nonlinear feedback}

On again combining (\ref{1.3}) and (\ref{1.4}),
we find the fundamental components,
%\begin{eqnarray}
%  2\left(\frac{\partial}{\partial T}+\textrm{i}Y\right)
%  \rho_1 &&+ m \cN p_1  \nonumber \\
%  =\varepsilon^2\delta^{-1} &&[-\mathrm{i}\cN(-\textrm{i}u_0w_1-v_1w_{0Y}
%  -\textrm{i}mw_0w_1-\delta^{-1}v_2w_{1Y}^*-v_1^*w_{2Y}+\textrm{i}mw_2w_1^*
%  -2\textrm{i}mw_1^*w_2)
%  \nonumber \\
%  &&-\textrm{i}u_0\rho_1-v_1\rho_{0Y}-\textrm{i}mw_0\rho_1
%  -\delta^{-1}v_2\rho_{1Y}^*-v_1^*\rho_{2Y}+\textrm{i}mw_2\rho_1^*
%  -2\textrm{i}mw_1^*\rho_2].  \label{3.19}
%\end{eqnarray}
\begin{equation}
  \left(\frac{\partial}{\partial T}+\textrm{i}Y\right) \rho_1 +\frac{1}{2} m \cN p_1
  =-\varepsilon^2\delta^{-1}\mathrm{i} u_0\rho_1, \label{3.23}
\end{equation}
with the leading-order nonlinear terms included on the right,
and after a considerable number of cancellations
stemming from the use of
(\ref{3.4}), (\ref{3.5}), (\ref{3.11}) and (\ref{3.12})
and the leading-order relations
$\rho_1=-\mathrm{i}\cN w_1$ and $v_{1Y}=-\mathrm{i}mw_1$.
Note that the nonlinear terms generated by the first harmonic
and mean-flow components $w_0$ and $\rho_0$ completely cancel out at this stage,
leaving only the effect of the modification to the streamwise mean flow $u_0$.
But the scaling established for the mean flow correction
implies that the right-hand side of (\ref{3.23}) is
$O(\delta^{-3}\varepsilon^2)$. Thus, {the mean flow
feedbacks on the fundamental} mode when $\delta=\varepsilon^{2/3}$.
That is, for
\begin{equation}
    t=O(\varepsilon^{-\frac{2}{3}}), \quad y=\cN+O(\varepsilon^{\frac{2}{3}}).
\end{equation}
These are the scalings for the nonlinear critical layer theory
outlined in the next section.

Note that we may extend the analysis to consider the higher harmomics.
One finds that when $\delta=\varepsilon^{2/3}$, the Fourier component $e^{\mathrm{i}j(x+mz)}$ with $j>1$
is $O(\varepsilon^{j/3})$, which signifies that the higher-order harmonics
$j\geq3$ are still weak when the mean flow correction
begins to feedback on the fundamental. Thus, they play no role in the
nonlinear theory.

%\begin{equation}
%  \textrm{i} \cN u_1 - (f-1)v_1 + \mathrm{i}p_1 = -
%  \varepsilon^2 \delta^{-1} v_1u_{0Y} \label{3.20}
%\end{equation}

\section{Nonlinear critical-layer theory}

\subsection{The reduction}

Motivated by the weakly nonlinear analysis, we now introduce
the rescalings,
\begin{equation}
  T=\varepsilon^{\frac{2}{3}}t, \quad Y=\frac{y-\cN}{\varepsilon^{\frac{2}{3}}}.
  \label{44}
\end{equation}
The outer solution for the pressure is
\begin{equation}
  p=\varepsilon p_1e^{\mathrm{i}(x+mz)}+\mathrm{c.c.},
  \qquad
  p_1=\left\{\begin{array}{ll}
   A(T)\hat{p}_{A}(y),
  &
  y>\cN,\\
 A(T)\hat{p}_{A}(y)+ B(T)\hat{p}_{B}(y), & 0< y<\cN,\end{array}\right.
\label{4.2}
\end{equation}
which is a single dominant Fourier mode characterized by the steady wave solution.
However, the amplitudes $A$ and $B$ now evolve with the slow time $T$,
because the nonlinear evolution of critical layer can affect the outer flow.
Initially, $A$ and $B$ are given by the linear analysis:
\begin{equation}
A(0)=\frac {A_{L}}{\varepsilon}, \quad B(0)=\frac{B_{L}}{\varepsilon}.
\end{equation}
%with $|A(0)|=1$ in view of (\ref{2.30}).

Inside the critical layers, we set
\begin{eqnarray}
  p=&&\varepsilon A(T)e^{\mathrm{i}(x+mz)}+\mathrm{c.c.}+...,
  \quad
  [w,\rho]=\varepsilon^{\frac{1}{3}}[w_1(Y,T),\rho_1(Y,T)] e^{\mathrm{i}(x+mz)}+\mathrm{c.c.}+...
      \nonumber \\
  &&
  [u,v]=\varepsilon [u_1(Y,T),v_1(Y,T)] e^{\mathrm{i}(x+mz)}+\mathrm{c.c.}
  +\varepsilon^{\frac{2}{3}}[U_0(Y,T),0]+...
\end{eqnarray}
Equation (\ref{3.23}) and the leading-order
fundamental-mode components of (\ref{1.1}), (\ref{1.3})-(\ref{1.5})
and (\ref{1.6}) now
become
\begin{equation}
  \frac{\partial \rho_1}{\partial T}+\textrm{i}Y\rho_1 +
  \frac{m\cN}{2}A=-\textrm{i}U_0\rho_1.  \label{4.7}
\end{equation}
\begin{equation}
\textrm{i}\cN u_1-(f-1)v_1+\textrm{i}A=-v_1U_{0Y},  \label{4.9}
\end{equation}
%\begin{equation}
%\pm\mathrm{i}\cN V_1+fU_1+P_{1Y}=0, \label{4.9a}
%\end{equation}
\begin{equation}
  w_1= \frac{\textrm{i}}{\cN}\rho_1,\qquad
  %\label{4.9b}
%\end{equation}
%\begin{equation}
\quad v_{1Y}=-\textrm{i}m w_1,  \label{4.9c}
\end{equation}
\begin{equation}
  \cN^2 u_{1Y}+\mathrm{i}m(f-1-U_{0Y})\rho_1=\mathrm{i}\cN v_1U_{0YY}.
  \label{4.10a}
\end{equation}
 The initial condition of $\rho_1$ is given by the linear result
\begin{equation}
  \rho_1\rightarrow - \frac{\mathrm{i}m\cN A(0)}{2}
  \frac{e^{-\mathrm{i}YT}-1}{Y}, \quad T\rightarrow 0.
\end{equation}
Similar to (\ref{3.4aa}), the mean-flow velocity $U_0$ is governed by
\begin{equation}
\frac{\partial U_{0}}{\partial T}=-v_1^*u_{1Y}+\textrm{i}mw_1u_1^*+\mathrm{c.c.}
  \label{4.10}
\end{equation}
The initial condition is $U_0\rightarrow 0$ as $T\rightarrow 0$, as in early linear evolution the mean-flow modification is minimal.

It is possible to algebraically manipulate (\ref{4.7})-(\ref{4.10a})
and then integrate in $T$ to show that
\begin{equation}
U_0=-\frac{2}{\cN^3}|\rho_1|^2, \label{4.10b}
\end{equation}
a result that can be traced back to the fact that
the change to the mean flow is given by
the Eulerian pseudo-momentum \citep{Buhler14}, which is the right-hand side of
(\ref{4.10b}) to leading order in the critical layer.
Hence
\begin{equation}
  \frac{\partial \rho_1}{\partial T}+\textrm{i}Y\rho_1
  +\frac{1}{2}m\cN A=  \textrm{i}\frac{2}{\cN^3}|\rho_1|^2\rho_1.
  \label{4.12}
\end{equation}

%The nonlinear term is a Landau cubic term with imaginary part; this is generic in nonlinear dynamics of non-dissipative systems with few degrees of freedom, for example such terms exist in nonlinear Schr\"{o}dinger equation, Stokes waves and Duffing oscillator.

To match the inner and outer solutions, we first note,
from (\ref{4.9c}), that $v_{1Y}= m\cN^{-1}\rho_1$. Integrating this relation in $Y$ over the critical layer then provides the jump of the outer solution $v_1={\textrm{i}(yp_{1,y}-fp_1)}/{[y^2-f(f-1)]}$ for the limit of $y\to\cN$ ({\it cf.} (\ref{2.11}$b$)), which yields
\begin{equation}
  B=-\textrm{i}m\frac{f(f-1)-\cN^2}{\cN^2}
  \int_{-\infty}^{\infty}\rho_1\textrm{d}Y,
  \label{4.16}
\end{equation}
in a similar manner to \S 3.2 and (\ref{2.26}).

Last, we again use the forcing condition at $y=0$ to close the problem:
\begin{align}
    (A-A^*)\hat{p}_{A}(0)+(B-B^*)\hat{p}_{B}(0) &=0, \nonumber \\
  (A+A^*)\hat{p}'_{A}(0)+(B+B^*)\hat{p}'_{B}(0) &=
  -f\frac{\varepsilon_0}{\varepsilon}
  \label{4.18}
\end{align}
({\it cf.} \S 3.3 and (\ref{forc1})). Note that the form of the forcing impacts the reduced model
      only through the closure relations in (\ref{4.18}). Had we used a
      different idealization of the forcing
      here, there would be a different algebraic
      relation between $A$, $B$ and $\varepsilon_0/\varepsilon$.
      However, this relation still connects $A$ with the
      forcing amplitude and the integral of $\rho_1$ over the
      critical layer, and in the scaled, canonical system presented below,
      all that would change would be how the parameters of that
      system (denoted $c_0$, $c_1$ and $c_2$ in \S5.2)
      depend on the original physical constants. In this sense, the
      reduced model is independent of the choice of forcing.

\subsection{Canonical system}

The final rescalings
\begin{equation}
  \rho_{1}=\left(\frac{m\cN^4}{4}\right)^{\frac{1}{3}}\gamma(\eta,\tau),\quad
  T=\left(\frac{2\cN}{m^2}\right)^{\frac{1}{3}}\tau,\quad
  Y=\left(\frac{m^2}{2\cN}\right)^{\frac{1}{3}}\eta,
\end{equation}
lead to the canonical form,
\begin{equation}
  \frac{\partial \gamma}{\partial \tau}+\mathrm{i}\eta \gamma+A=\mathrm{i}
  | \gamma|^2 \gamma,  \label{4.22}
\end{equation}
\begin{equation}
  A(\tau)=c_0 +
 \frac{\mathrm{i}c_1}{\pi}\int_{-\infty}^{\infty}\gamma_r \mathrm{d}\eta
 -\frac{c_2}{\pi}\int_{-\infty}^{\infty}\gamma_i\mathrm{d}\eta,
 \label{4.23}
\end{equation}
where $\gamma=\gamma_r+\mathrm{i}\gamma_i$,
%\begin{equation}
%  c_1=\frac{m^2[f(f-1)-\cN^2]}{2\cN}\frac{\hat{p}_{B}(0)}{\hat{p}_{A}(0)},\quad
%  c_2=\frac{m^2[f(f-1)-\cN^2]}{2\cN}\frac{\hat{p}'_{B}(0)}{\hat{p}'_{A}(0)}.
%  \label{4.21}
%\end{equation}
\begin{equation}
  c_0 = -\mathrm{sgn}\left(\frac{f}{\hat{p}_{A}'(0)}\right)
  \frac{|1+c_1c_2|}{\sqrt{1+c_1^2}},
\end{equation}
and
\begin{equation}
  \left\{ \begin{array}{r} c_1 \cr c_2 \end{array} \right\}
  = \frac{\pi m^2[f(f-1)-\cN^2]}{2\cN} \left\{\begin{array}{l}
  {\hat{p}_{B}(0)}/{\hat{p}_{A}(0)} \cr {\hat{p}'_{B}(0)}/{\hat{p}'_{A}(0)}
  \end{array} \right\} .
  \label{4.21}
\end{equation}
For $\tau\ll1$, we must match $\gamma(\eta,\tau)$
to the corresponding solution of the linear problem, given by
\begin{equation}
  \gamma = \mathrm{i}A\frac{1-e^{-\mathrm{i}\eta\tau}}{\eta},
  \qquad
  A =
%  -\mathrm{sgn}\left[\frac{f}{\hat{p}_{A}'(0)(1+c_1c_2)}\right]
  \frac{ c_0 (1-\mathrm{i} c_1)}{1+c_1c_2} ,
  % \quad \tau\rightarrow 0.
  \label{4.25}
\end{equation}
which provides the initial condition for (\ref{4.22}).

The reduced model equations in (\ref{4.22})-(\ref{4.25})
are solved numerically in the next section. The system
is integro-differential in the sense that (\ref{4.22})
is an equation of motion in time, solved at each level of $\eta$,
with the integral constraint in (\ref{4.23}). There is no
dependence on either $x$ or $z$, because the leading-order dynamics involves only the fundamental mode of the forcing wave pattern
and the mean-flow response (which is then prescribed by the
pseudo-momentum). The only
nonlinearity is the cubic term on the right of (\ref{4.22}),
which is generic in weakly nonlinear theories
of non-dissipative systems with few degrees of freedom.
The model is therefore rather different from those that
emerge for classical forced critical layers, which usually
take the form of partial differential equations
in all the spatial variables. The reduced model
has the two parameters, $c_1$ and $c_2$,
and the choice of sign for $f\hat{p}'_A(0)$ in $c_0$.
In most situations $\hat{p}_{A}$ and
$\hat{p}_{B}$ are characterized by a similar exponential away from $y=0$,
implying $c_1\approx c_2$.

From (\ref{4.22})-(\ref{4.23}), one can establish that
the quantity,
\begin{equation}
  {\cal H}=
  \int_{-\infty}^{\infty}\left[\frac{1}{2}|\gamma|^4-\eta|\gamma|^2+
    2\mathrm{Im}(A^*\gamma)\right]{\mathrm{d}\eta}
  +\frac{c_1}{\pi}\left[\int_{-\infty}^\infty\gamma_r{\mathrm{d}\eta}
    \right]^2+\frac{c_2}{\pi}\left[\int_{-\infty}^\infty\gamma_i
    {\mathrm{d}\eta}\right]^2,
  %\equiv\frac{ c_1(1+c_1c_2)}{1+c_1^2}.
  \label{claw}
\end{equation}
must be conserved, and therefore equal to $\pi c_1(1+c_1c_2)/(1+c_1^2)$
in view of the initial conditions.
This constraint implies that
the linear-in-time growth of $\gamma(\eta,\tau)$ predicted by linear theory
must eventually become arrested, as otherwise the quartic
first term in (\ref{claw}) cannot be counter balanced by the remaining
quadratic and constant terms.
To determine the manner in which the arrest takes place, we turn
to a numerical solution of the reduced model.

\subsection{Numerical solutions}

To solve the canonical system of equations numerically,
we first select a grid in
$\eta$ spanning a finite domain (we use 1501 equally spaced gridpoints
over the interval $1.5<\eta<3$ where $\gamma$ has large gradients,
then 1544 gridpoints distributed evenly
over $-25<\eta<1.5$ and $3<\eta<25$). We then
integrate (\ref{4.22}) forward in time numerically using a 4th-order
Runge-Kutta method at each of the grid points. To evaluate the integrals
in (\ref{4.23}), we use an approach similar to Warn \& Warn (1978) to
extrapolate the limits to infinity. We use parameter settings
guided by the computations of of Marcus \textit{et al.} (2013):
$m=1/2$, $f=4/3$, $\cN=4/3$, which yield $c_1=0.238$, $c_2=0.219$.

\begin{figure}
\begin{center}
   \includegraphics[width=0.85\linewidth,height=0.3\linewidth]{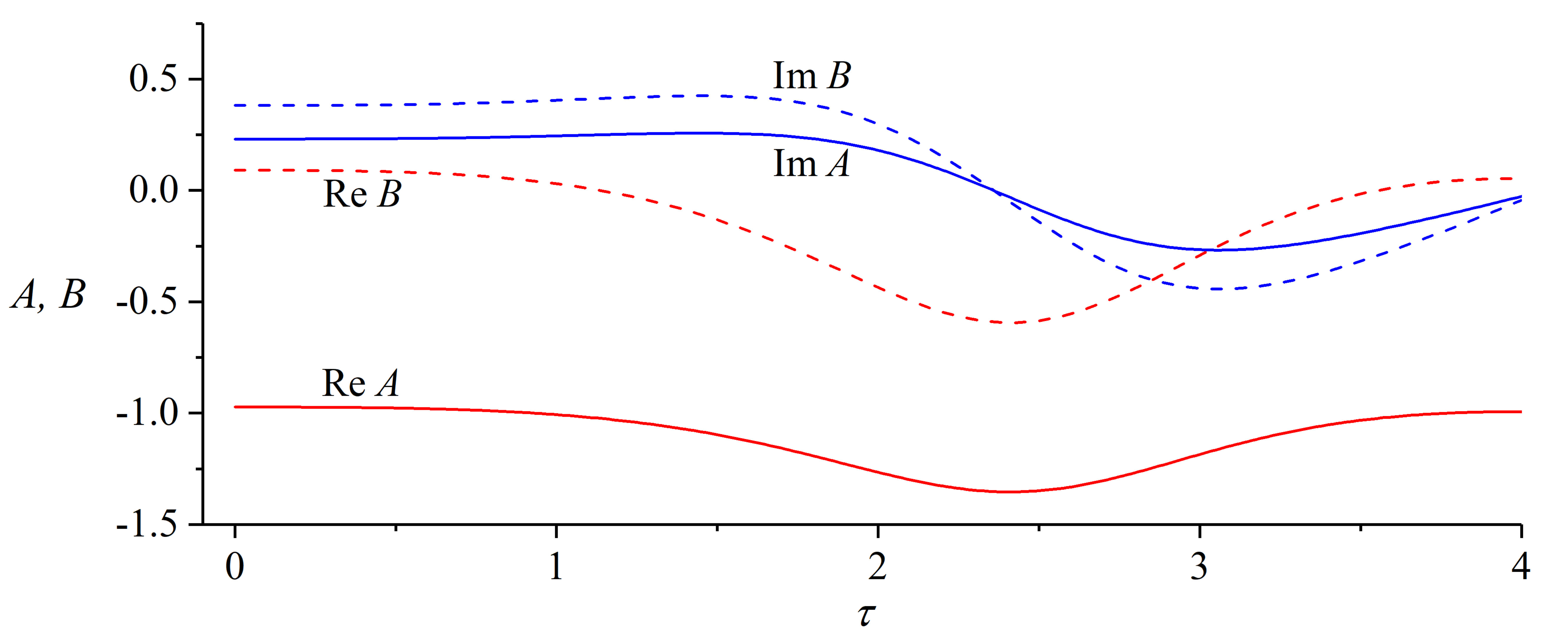}
   \caption{Evolution of $A$ and $B$ with $\tau$; $m=1/2$, $f=4/3$, $\cN=4/3$.
}
   \label{F3}
 \end{center}
\end{figure}

Figure \ref{F3} displays the evolution in $\tau$
of the forced wave amplitudes, $A$ and $B$, which is relatively mild
with Re$(A)\approx c_0\approx -1$ and
Im$(A)$, Re$(B)$ and Im$(B)$ all remaining small. This mild behaviour
results because, in (\ref{4.23}), $|c_1|$ and $|c_2|$
are fairly small. Thus, the forced wave evolves slowly
over the bulk of the shear flow
({\it i.e.} the outer region), maintaining a profile similar
to the linear distribution in figure \ref{F2}.

%\begin{figure}
%\begin{center}
% \includegraphics[width=0.497\linewidth]{gamma_r.png}
% \includegraphics[width=0.497\linewidth]{gamma_i.png}
% \caption{\textcolor[rgb]{0.00,0.07,1.00}{Evolution of $\gamma(\eta,\tau)$; the two dashed lines represent the linear theory's prediction of the largest value of Re$(\gamma)$ and Im$(\gamma)$, respectively; }
%   $m=1/2$, $f=4/3$, $\cN=4/3$.}
%   \label{F4}
% \end{center}
%\end{figure}

%\begin{figure}
%\begin{center}
% \includegraphics[width=0.495\linewidth]{gamma_tau=1.png}
% \includegraphics[width=0.495\linewidth]{gamma_tau=3_5.png}
% \caption{\textcolor[rgb]{0.00,0.07,1.00}{Snapshots of $\gamma(\eta,\tau)$ at $\tau=1$ and $\tau=3.5$; the thick and thin solid line represents the numerical solution's real and imaginary part, respectively, and the dashed and dot line represents the linear theory's prediction of the real and imaginary part, respectively; }
%   $m=1/2$, $f=4/3$, $\cN=4/3$.}
%   \label{F4a}
% \end{center}
%\end{figure}

\begin{figure}
\begin{center}
  \includegraphics[width=0.85\linewidth]{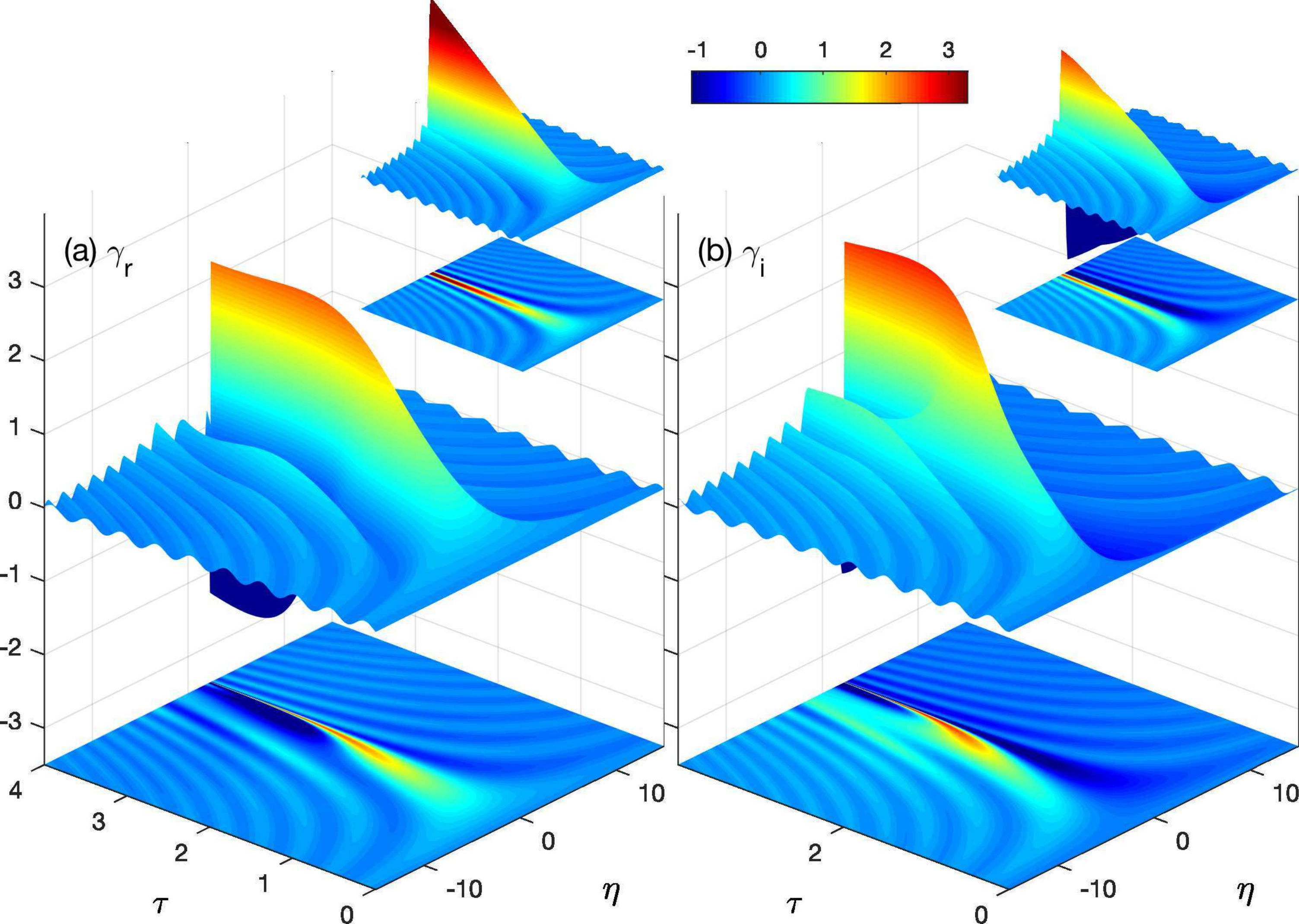}
  \end{center}
  \vspace{0.25cm}
\begin{center}
 \includegraphics[width=0.75\linewidth]{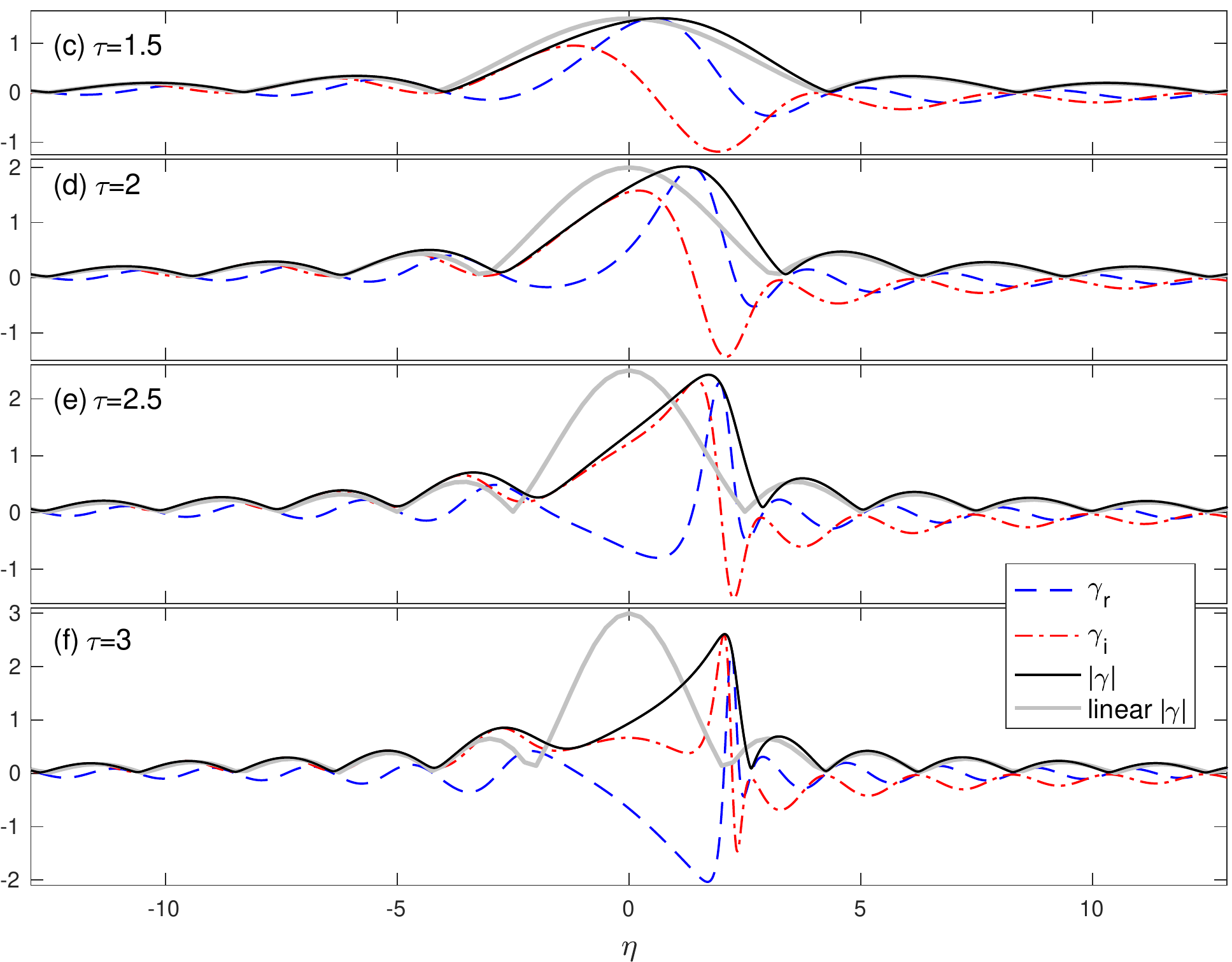}
 \end{center}
 \caption{
   (a) Real and (b) imaginary parts of the critical-layer density perturbation
   $\gamma(\eta,\tau)$, shown as surfaces above
   the $(\eta,\tau)-$plane. To prevent the viewing perspective
   from obscuring parts of the solution, we also show
   density maps of the solutions underneath.
   The insets show corresponding plots of the linear solution in (\ref{4.25}).
   Panels (c)--(f) plot snapshots of $\gamma(\eta,\tau)$ at the times
   indicated; the linear result for $|\gamma(\eta,\tau)|$ is also included.
   ($m=1/2$, $f=4/3$, $\cN=4/3$.)}
   \label{F4}
\end{figure}

The density perturbation $\gamma(\eta,\tau)$,
shown in figure \ref{F4}, exhibits a richer behaviour:
for $\tau<1$, the numerical solution follows the linear prediction
in (\ref{4.25}), with its characteristically developping undulations
and linear growth near $\eta=0$ (see figure \ref{F4}(a,b)). Once $|\zeta|$
reaches order-one values there, however, the growth of the numerical solution
saturates, as demanded by the constraint in (\ref{claw}).
Despite this, the solution continues to undulate over increasingly shorter
spatial scales. Moreover, nonlinear
effects distort the density profile further, shifting the maximum
magnitude from $\eta=0$ to a small, positive level in $\eta$
and generating pronounced fine structure over a narrow region
nearby.

\begin{figure}
\begin{center}
 \includegraphics[width=0.45\linewidth]{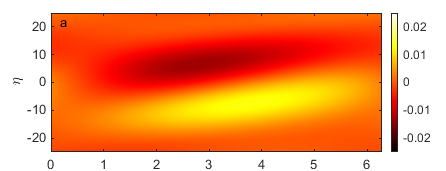}
 \includegraphics[width=0.4\linewidth]{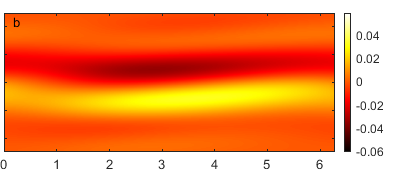}
 \includegraphics[width=0.45\linewidth]{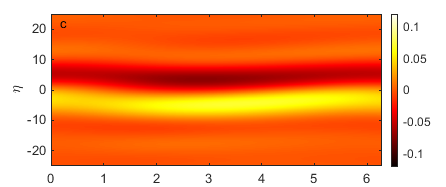}
 \includegraphics[width=0.4\linewidth]{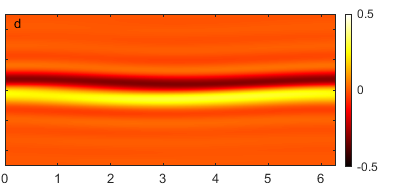}
 \includegraphics[width=0.45\linewidth]{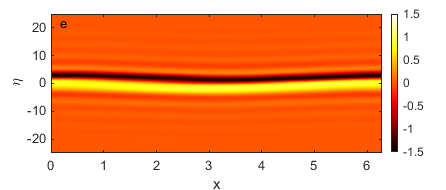}
 \includegraphics[width=0.415\linewidth]{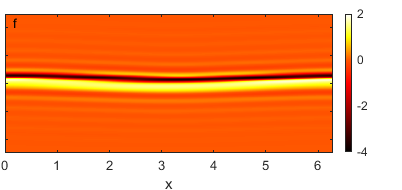}
 \caption{
   Snapshots of
   vertical vorticity $\zeta$ within the baroclinic critical layer
   near $y=\cN=4/3$, plotted as a colormap on the $(x,\eta)-$plane
   for (a) $\tau=0.3$, (b) $\tau=0.45$, (c) $\tau=0.6$, (d) $\tau=1$,
   (e) $\tau=1.5$, and (f) $\tau=1.8$.
      The domain plotted is $|\eta|<25$,
   corresponding to $|y-\cN|<0.39$, at cross-section $z=0$
   and over one streamwise wavelengths of the forcing pattern.
   ($\varepsilon_0=0.05$, $\varepsilon=0.0062$, $m=1/2$,
   $f=4/3$, $\cN=4/3$.)
   }
   \label{F5}
 \end{center}
\end{figure}

The rapid spatial variation in $\gamma(\eta,\tau)$
significantly impacts the critical-layer vorticity, which depends on
the $\eta-$derivatives of $\gamma(\eta,\tau)$.
In particular, the leading-order vertical vorticity
{is given by the mean-flow vorticity $\zeta_0$}:
\begin{equation}
\zeta \sim \zeta_0 \equiv \frac{\partial}{\partial\eta} |\gamma|^2.
\label{5.2}
\end{equation}
However, from the matched asymptotic expansion, we may reconstruct
$\zeta(x,\eta,z,t)$ to higher orders,
{incorporating the fundamental Fourier mode $\zeta_1$ and
  first harmonic $\zeta_2$}, as summarized in Appendix \ref{appo}.
The evolution of the reconstructed vertical vorticity field is plotted in
figure \ref{F5}. For early times, $\zeta_0\ll1$, and the vertical vorticity
is actually given by the higher-order linear solution
%\begin{equation}
%  \zeta \sim \left(\frac{16\varepsilon m^4}{\cN^2}\right)^{\frac{1}{3}}(f-1)
%  \eta^{-1}\sin \frac{\eta\tau}{2}\sin\left(x+mz-\frac{1}{2}\eta\tau
%  +\arg A(0)\right)
%\end{equation}
(as in figure \ref{F5}a, \textit{cf} (\ref{2.21})).
With the increase of $\tau$, the vorticity distribution tilts over
and $\zeta_0$ grows to dominate $\zeta$, as seen in figure
\ref{F5}b,c. This growth leads to
the distinctive dipolar stripe seen in figure \ref{F5}d.
In the later stages of evolution (figure \ref{F5}e,f), the
stripe becomes stronger and more focussed, shifting slightly above $\eta=0$,
and corresponding to the sharpening oscillations in $\gamma$ seen in
figure \ref{F4}.

The behaviour of the numerical solution seen in figures \ref{F3}-\ref{F5}
is generic for most parameter settings;
%based on the configuration
%sketched in figure \ref{F 1}:
for moderate $m$, $f$ (either $f>1$ or $f<0$) and $\cN$, the parameters
$c_1$ and $c_2$ of the reduced model are relatively small in magnitude,
prompting similar dynamics.
Even when $|c_1|$ and $|c_2|$ become order one, the evolution still bears
qualitative similarities. However,
more complicated behaviour can occur in the reduced model
when these parameters take higher values. Such parameter settings
can be achieved at special combinations of $m$, $f$ and $\cN$ for which
$\hat{p}_A(0)$ becomes small, or perhaps for other types of forcing.
We avoid consideration of
special situations of this sort,
and instead turn to a deeper analysis of the focussing dynamics observed
in the reduced model.

\subsection{Long-time focussing}

%Guided by our observations of the numerical solutions above
%we now consider the limit $(c_1,c_2)\to0$.
%In this situation, $A=-$sgn$(f \hat{p}_A'(0))$ remains constant (as in
%the analytical special case of the forced Rossby wave critical layer
%discussed by Stewartson 1978), with
%the example considered above corresponding to $A=-1$.
 In view of the result that $A$ changes slowly, we now use the approximation of $A=constant$ to gain further analytical insights to the focussing phenomenon. This device was used previously by Stewartson (1978) to obtain an analytical solution to the nonlinear evolution of Rossby wave critical layers. In our model, constant $A$ in (\ref{4.23}) requires $c_1=c_2=0$, hence $A=-$sgn$(f \hat{p}_A'(0))$, which is $-1$ for the current parameter setting.
 The evolution equation
(\ref{4.22}) can then be written as the one-degree-of-freedom
Hamiltonian system,
\begin{eqnarray}
 \frac{\partial\gamma_r}{\partial\tau} = \frac{\partial H}{\partial \gamma_i} =
  1+\eta\gamma_i-\gamma_r^2\gamma_i-\gamma_i^3,\nonumber\\
 \frac{\partial\gamma_i}{\partial\tau} =-\frac{\partial H}{\partial \gamma_r}=
  -\eta\gamma_r+\gamma_r^3+\gamma_r\gamma_i^2, \label{5.9}
\end{eqnarray}
with Hamiltonian,
\begin{equation}
  H=-\frac{1}{4}\left(\gamma_r^2+\gamma_i^2\right)^2+
  \frac{1}{2}\eta\left(\gamma_r^2+\gamma_i^2\right)+\gamma_i
\end{equation}
(the point-wise version of the conserved quantity $\cal H$ in (\ref{claw})
for $c_1=c_2=0$).
For the specific initial condition of our critical-layer problem,
$H=0$ for all values of $\eta$.

\begin{figure}
\begin{center}
    \includegraphics[width=0.9\linewidth]{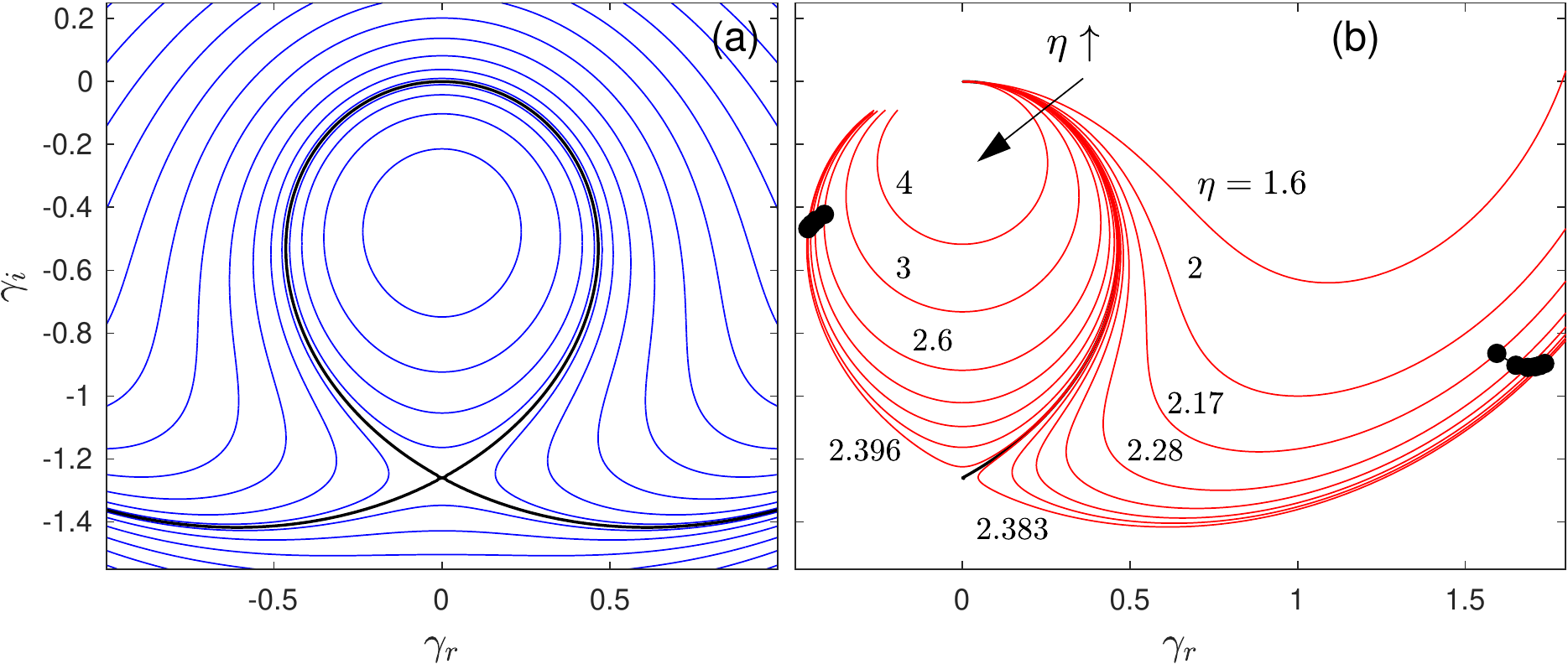}
   \includegraphics[width=0.8\linewidth]{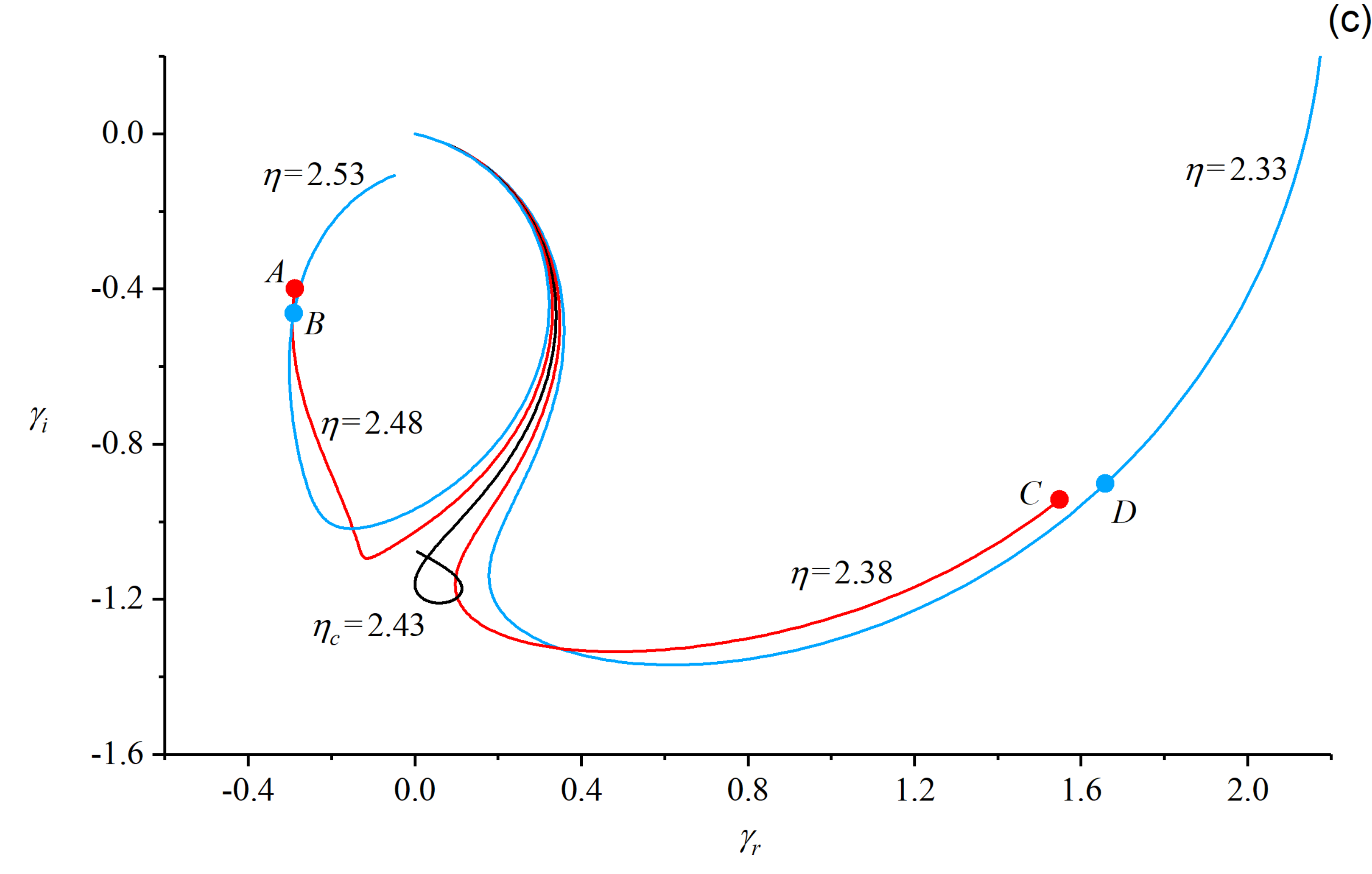}
   \caption{{Phase portraits of $(\gamma_r,\gamma_i)$ for
     (a) the Hamiltonian system (\ref{5.9}) with $\eta=3/\sqrt[3]{2}$
       and various $H$, with the thicker line indicating $H=0$},
     (b) trajectories from the point
     $(\gamma_r,\gamma_i)=(0,0)$ for a selection of values of $\eta$,
     and (c) the numerical solution of \S 5.3, at the five values of
     $\eta$ indicated. The black points in (b) and the
     (red and blue) pairs marked $(A,B)$ and $(C,D)$ in (c)
     have the same values of $|\eta-\eta_c| e^{\sigma \tau}$.
     %(with
     %$(\tau,\eta-\eta_c)=(3.5,0.05)$ for $A$, $(3,0.1)$ for $B$,
     %$(3.5,-0.05)$ for $C$, and $(3,-0.1)$ for $D$).
   }
   \label{F9}
 \end{center}
\end{figure}

Figure \ref{F9}(a) illustrates the phase
portrait of the system (\ref{5.9})
for the special choice $\eta=\eta_c=3/\sqrt[3]{2}$.
In this case, the orbit from $(\gamma_r,\gamma_i)=(0,0)$
lies along a separatrix that converges to
a saddle point at $(\gamma_r,\gamma_i)=(0,-\gamma_e)$,
for $\tau\to\infty$, with $\gamma_e=\sqrt[3]{2}\approx1.26$. Trajectories from
$(\gamma_r,\gamma_i)=(0,0)$ for a spread of values of $\eta$ around
$\eta_c$ are illustrated in figure \ref{F9}(b);
the presence of the separatrix at $\eta=\eta_c$
implies that these trajectories bifurcate in direction on the phase plane
on passing through that special level. Thus,
a small variation in $\eta$ about $\eta_c$
can result in a large change of $\gamma$ at later
times, implying high values of $\gamma_\eta$ to feed into $\zeta$.

For the numerical solutions of \S 5.3, although $c_1$ and $c_2$
do not vanish, the forced-wave amplitude does
remain slowly varying in $\tau$, leading to
a qualitatively similar dynamics:
figure \ref{F9}(c) plots the phase portrait of $\gamma$
for five values of $\eta$ within the region where the dipolar
stripe is focussed. As $\eta$ varies from 2.38 to 2.48, the trajectories
for different levels abruptly switch in direction
near the point $(\gamma_r,\gamma_i)=(0,-1.2)$. Although the slow
variation of $A(\tau)$ precludes any trajectory from reaching
a steady value, the numerical solution for $\eta=2.43$
slows down, lingers and hesitates before selecting
one of the two possible directions, much like the
orbits for $c_1=c_2=0$ near the separatix in figure \ref{F9}(a,b).
The level of this trajectory
is slightly shifted from $3/\sqrt[3]{2}\approx 2.38$
because $c_1$ and $c_2$ are non-zero and $A(\tau)\ne-1$.
Nevertheless, we conclude that the close passage to an effective
saddle point on the $(\gamma_r,\gamma_i)$ phase plane is responsible
for the focussing effect. For the numerical solution, we therefore
define $\eta=\eta_c\approx2.43$ to be the level for which $\gamma$ evolves
slowest near the effective saddle, and refer to this location
as the nonlinear critical level.

Continuing the analysis for $c_1=c_2=0$,
we may linearize the system (\ref{5.9}) about
$\gamma=-\mathrm{i}\gamma_e$ to find that
\begin{equation}
    \frac{\partial}{\partial\tau}
    \left(\begin{array}{c} \gamma_r \cr \gamma_i+\gamma_e
  \end{array} \right)
    =
    \left( \begin{array}{cc}
    0 &  \eta-3\gamma_e^2  \cr
     -\eta+\gamma_e^2  & 0
  \end{array} \right)
    \left(\begin{array}{c} \gamma_r \cr \gamma_i+\gamma_e
  \end{array} \right)
. \label{5.14}
\end{equation}
The two eigenvalues of the matrix are $\pm\sigma$, with
corresponding eigenvectors $\mathbf{v}_+$ and $\mathbf{v}_-$, where
\begin{equation}
  \sigma=\sqrt{\left[(\gamma_e^2-\eta)\right]
    \left[\eta-3\gamma_e^2)\right]}
  \approx \sigma_c = \frac{\sqrt{3}}{\sqrt[3]{2}}
  \quad {\rm if} \quad \eta\approx\eta_c .
  \label{5.15}
\end{equation}
The solution of (\ref{5.14}) is then
\begin{equation}
    \left(\begin{array}{c} \gamma_r \cr \gamma_i+\gamma_e
  \end{array} \right)
    =r_+\mathcal{\mathbf{v}}_+e^{\sigma(\tau-\tau_0)}
    +r_-\mathcal{\textbf{v}}_-e^{-\sigma(\tau-\tau_0)},  \label{5.17}
\end{equation}
for some constants $r_\pm$ and a time constant $\tau_0$ indicating
when the orbit reaches the neighbourhood of the saddle point.

\begin{figure}
\begin{center}
\includegraphics[width=0.493\linewidth]{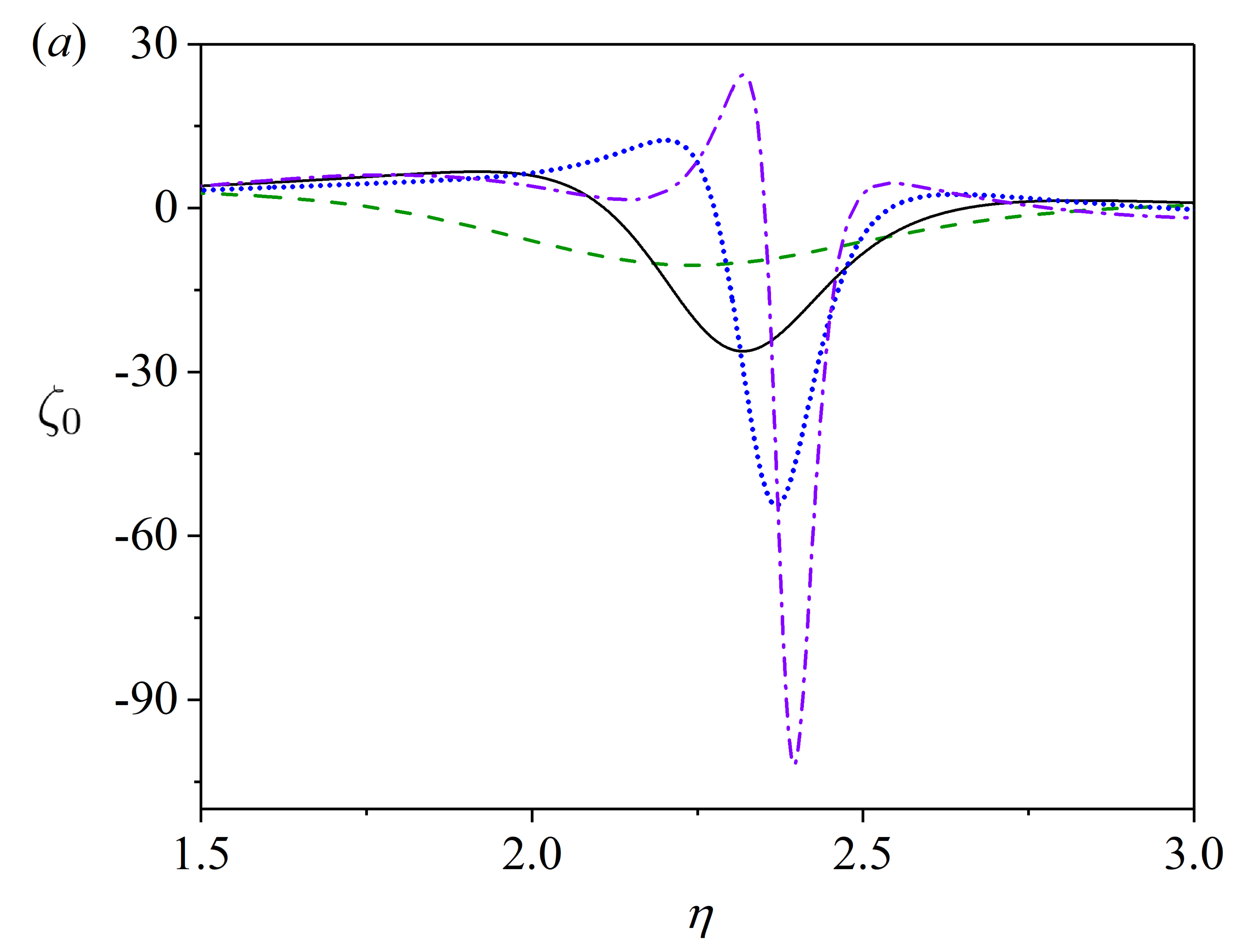}
 \includegraphics[width=0.494\linewidth]{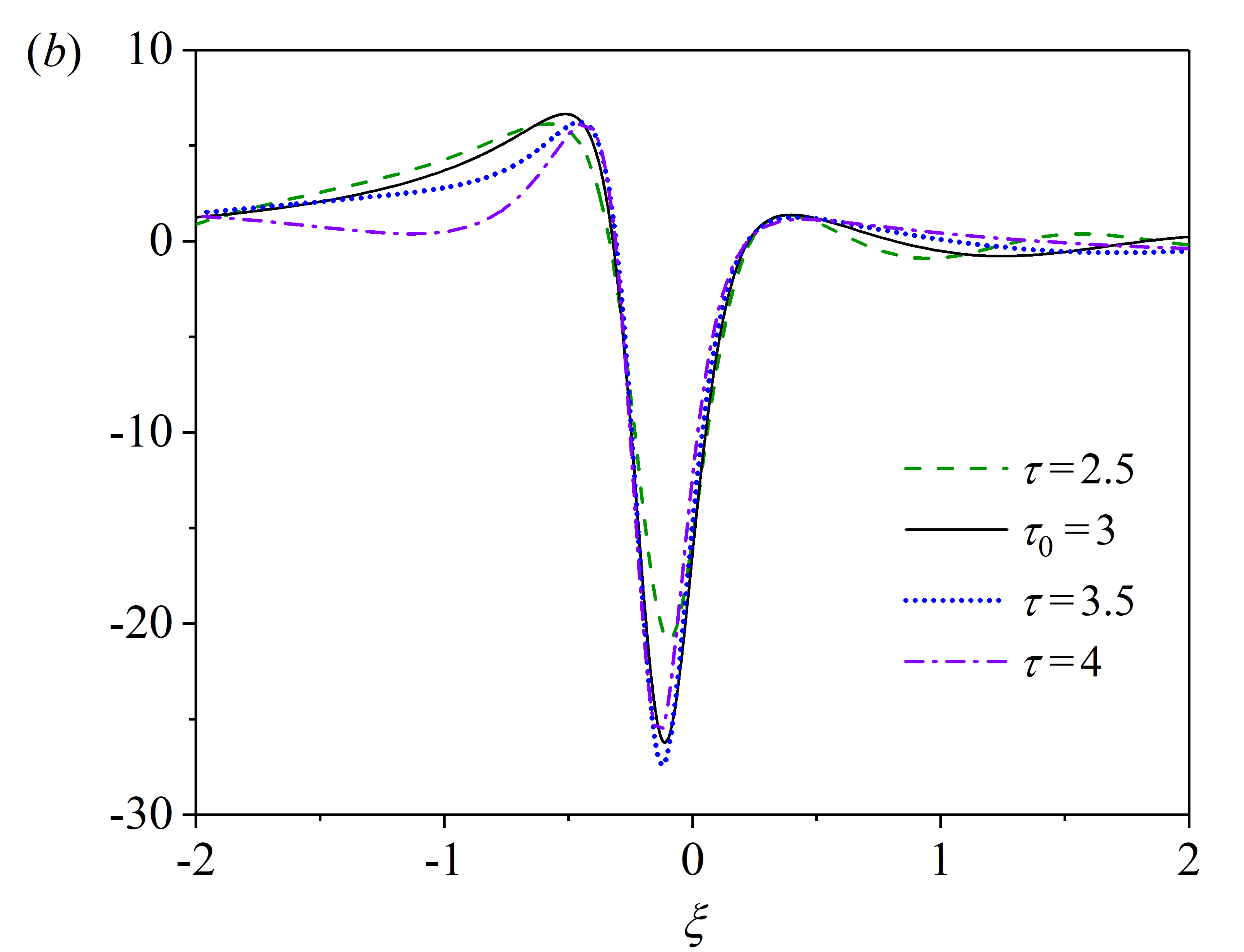}
 \caption{(a) Evolution of $\zeta_0$ near $\eta_c=2.43$ at the times
   indicated. (b) Scaled profiles,
   $\zeta_0 e^{-\sigma_c(\tau-\tau_0)}$ against
   $\xi=(\eta-\eta_c) e^{\sigma_c(\tau-\tau_0)}$, choosing $\tau_0=3$;
   ($m=1/2$, $f=4/3$, $\cN=4/3$).}
   \label{F6}
 \end{center}
\end{figure}

Now, along the separatrix converging to
$\gamma=-\mathrm{i}\gamma_e$ for $\eta=\eta_c$,
the constant $r_+$ must vanish. But
when $\eta$ is close to, but not at $\eta_c$, this factor
is small but finite, hence a local linearization of $r_+(\eta)$ near $\eta=\eta_c$ leads us to set
$r_+\approx C (\eta-\eta_c)$, for some constant $C$. Therefore,
\begin{equation}
    \left(\begin{array}{c} \gamma_r \cr \gamma_i+\gamma_e
    \end{array} \right) \sim C (\eta-\eta_c) \mathcal{\mathbf{v}}_+
    e^{\sigma_c(\tau-\tau_0)}
    ,
\end{equation}
at large times.
That is, for $\eta$ near $\eta_c$, those pairs of $(\eta,\tau)$
with the same $(\eta-\eta_c) e^{\sigma_c\tau}$ should have the same $\gamma$.
Although this property is derived from the local linearization
about the fixed point, it still holds when trajectories have
progressed further along the unstable manifolds of that saddle
because the trajectories shadow that curve.
This is illustrated in figure \ref{F9} for both the Hamiltonian system
and the numerical solution,
where the pairs of points plotted along sample orbits
have the same values
for $(\eta-\eta_c) e^{\sigma_c\tau}$, and therefore similar $\gamma$,
even though they correspond to different choices of $(\eta,\tau)$.
We can express the property mathematically
by writing the solutions in the self-similar form,
\begin{equation}
  \gamma \approx F(\xi) \quad {\rm and} \qquad
  \zeta_0 \approx e^{\sigma_c(\tau-\tau_0)} \frac{\rm d}{{\rm d}\xi} |F(\xi)|^2
  , \quad {\rm with} \quad
  \xi=(\eta-\eta_c) e^{\sigma_c(\tau-\tau_0)}, \label{5.18}
\end{equation}
for some function $F(\xi)$ related to the shape of the unstable
manifolds of the saddle point.
Thus, the lengthscale of the nonlinear critical layer at $\eta=\eta_c$
decreases exponentially in time, accounting for the relatively
rapid focussing of sharp spatial variations in
$\gamma$ at later times in figure \ref{F4}, and the amplitude of
the vertical vorticity grows exponentially.
Figure \ref{F6} presents four snapshots of $\zeta_0(\eta,\tau)$
for the numerical solution, then replots them against $\xi$
and scaled by $e^{\sigma_c(\tau-\tau_0)}$, adopting $\tau_0=3$; while the profile of $\zeta_0$ keeps sharpening and strengthening, the rescaled profile remains nearly unchanged, confirming the self-similar structure in (\ref{5.18}).
%the self-similar structure to the data is visible by the
%collapse at the core of the vorticity profiles.

The exponential focussing towards the nonlinear critical level is
problematic as it implies that the higher-order harmonics of the forcing
pattern, which are neglected in our nonlinear critical layer model,
grow faster than the re-arrangments of the mean flow. In particular,
one can deduce that the vertical vorticity of the $j^{th}$ Fourier component,
$\exp[j(\mathrm{i}x+\mathrm{i}mz)]$, grows like $e^{(j+1)\sigma_c\tau}$.
The model therefore fails once the solution becomes overly focussed,
heralding the onset of a further, more complicated, stage of
evolution.

%\textcolor[rgb]{0.00,0.59,0.00}{Note that in an even longer time, the effect of the variation of $A$ becomes important: its periodic oscillation generates multiple levels of focusing (see the left panels of figure \ref{sols}(a),(b),(c)). But we will not further discuss this as the model already fails. }

\section{Effects of diffusion}

The increasingly fine scales encountered in the critical layer
due to the exponential focussing suggest that dissipation may also
become prominent over later times, even if
small initially. To explore this possibility in more detail,
we return to the governing equations and include the
viscous terms $\nu \nabla^2(u,v,w)$ in (\ref{1.1})--(\ref{1.3})
and diffusive term $\kappa \nabla^2 \rho$ in (\ref{1.4}). We then take the
distinguished limit $(\nu,\kappa)=O(\varepsilon^2)$, which corresponds to the order
when dissipation first becomes important. In particular, with this sacling of
$\nu$ and $\kappa$, the
 dissipative terms are too small to affect the
 quasi-steady wave in the bulk of the flow,
but enter the analysis of the baroclinic critical layers owing to the
reduced spatial scale in $y$. Equation (\ref{4.7}) is now replaced with
\be
  \frac{\partial \rho_1}{\partial T}+\textrm{i}Y\rho_1 +
  \frac{m\cN}{2}A=-\textrm{i}U_0\rho_1 + \frac{(\nu+\kappa)}{2\ve^2}
  \frac{\partial^2\rho_1}{\partial Y^2}.
  \label{x.1}
\ee
%\textcolor[rgb]{0.00,0.59,0.00}{One can easily derive this modified equation along the line of \S 4 and \S 5 adding dissipative terms, but one also finds that at order $(\nu,\kappa)=O(\varepsilon^2)$, dissipative terms remain small in equations without time derivatives, \textit{i.e.} (\ref{3.4})-(\ref{3.12}) and (\ref{4.9})-(\ref{4.10a}).}
The Eulerian pseudomomentum is no longer equal to the mean-flow response, as in (\ref{4.10b}), and we have to return to the mean-flow evolution equation:
\be
\frac{\partial U_0}{\partial T}=\frac{m}{\cN^2}(A^*\rho_1+A\rho_1^*)
+ \frac{\nu}{\ve^2}   \frac{\partial^2U_0}{\partial Y^2}.
\label{x.2}
\ee
(following from the substitution of (\ref{4.9})-(\ref{4.10a})
into the modified version of
(\ref{4.10})).
The initial condition is still given by (5.20), the dissipative terms
being negligible at early times when the spatial scales are larger.
The closure relations given by the match to the
outer solution remain (\ref{4.16}) and (\ref{4.18}).
Equations (\ref{x.1}) and (\ref{x.2}) can be combined to furnish the
integral relation,
\be
\frac{\rm d}{{\rm d}T}
\int_{-\infty}^\infty \left( |\rho_1|^2 + \half \cN^3 U_0 \right) {\rm d}Y =
- \frac{(\nu+\kappa)}{\ve^2}
\int_{-\infty}^\infty \left|\frac{\partial\rho_1}{\partial Y} \right|^2 {\rm d}Y
,
\label{x.R}
\ee
provided that $\rho_1$ and $U_0$ decay sufficiently quickly for $|Y|\to\infty$.
We now briefly discuss the dynamics captured by this dissipative
version of the model, focussing on
the astrophysically relevant limit $\nu\ll\kappa$.

\subsection{Modified canonical system}

A scaling similar to that in \S 5.2, now furnishes the modified
canonical system,
\be
  \frac{\partial \gamma}{\partial \tau}+\mathrm{i}\eta \gamma + A = -
  \mathrm{i} \gamma \cU  + \lambda \frac{\partial^2\gamma}{\partial\eta^2},
  \qquad
  \frac{\partial\cU}{\partial\tau} = A^*\gamma+A\gamma^*
  \label{x.3}
\ee
and (\ref{4.23}),
where
\be
\cU(\eta,\tau) = \left(\frac{2\cN}{m^2}\right)^{\frac{1}{3}} U_0
\qquad {\rm and} \qquad
\lambda = \frac{\kappa\cN}{m^2\ve^2}.
  \label{x.4}
\ee
This system may be solved numerically. For the task, we now use a
Crank-Nicolson method to evolve the system in time and centred
finite differences method to evaluate spatial derivatives,
exploiting Newton iteration at each time step to solve the nonlinear equations.

\begin{figure}
\begin{center}
\includegraphics[width=0.9\linewidth]{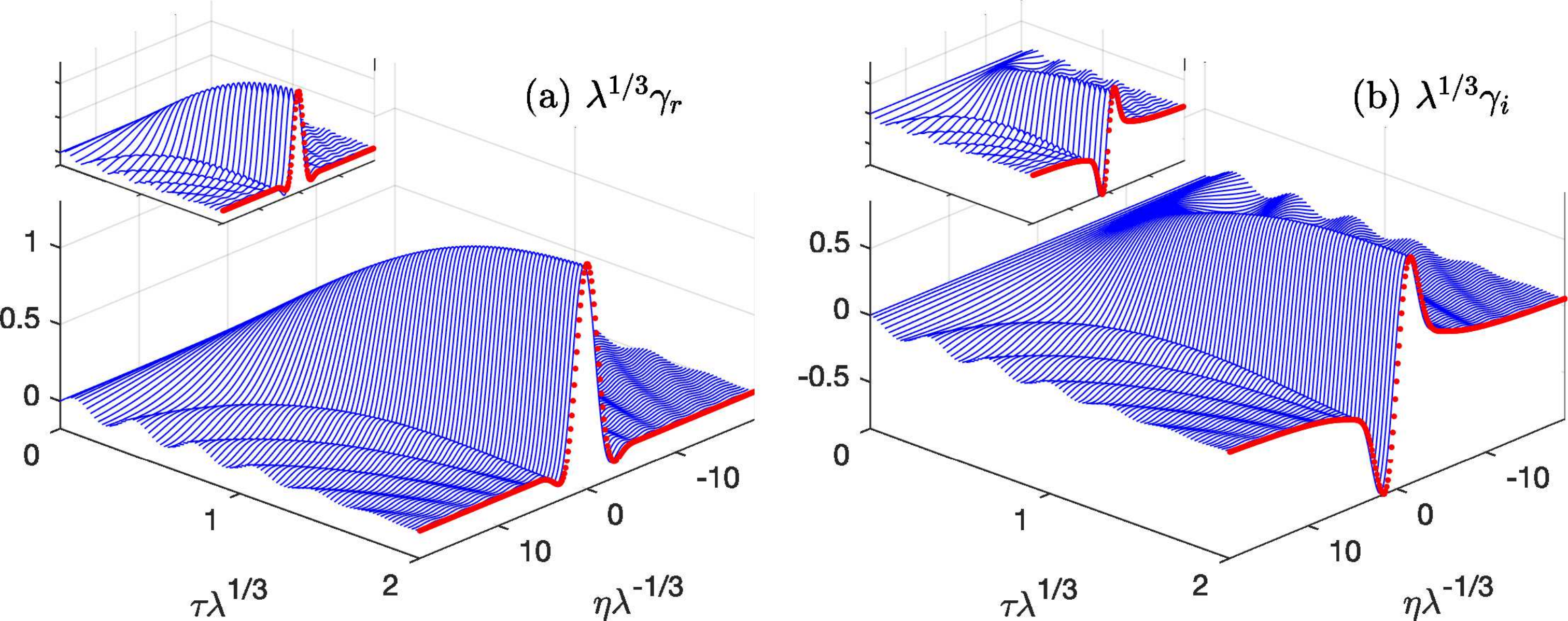}
\caption{
  The analytical solution (\ref{xxx.3}) for strong diffusion
  and $A=-1$,
  showing (a) $\lambda^{1/3}\gamma_r$ and (b) $\lambda^{1/3}\gamma_i$
  against the scaled space and time variables
  $\lambda^{-1/3}\eta$ and $\lambda^{1/3}\tau$. The (red) dots show the
  final steady-state solution. The insets show corresponding
  numerical solutions to the reduced model, computed for
  $\lambda=5.3$.
}
   \label{diffo}
 \end{center}
\end{figure}

Before characterizing the features of the numerical solutions,
we first pause to examine the dynamics in the limit that
diffusion is relatively strong, $\lambda\gg1$. In this limit,
the large diffusive term $\lambda\gamma_{\eta\eta}$ in (\ref{x.3})
must be balanced by introducing the rescalings,
$(\gamma,\tau)=O(\lambda^{-1/3})$, $\eta=O(\lambda^{1/3})$ and
$\cU=O(\lambda^{-2/3})$. The advection of the density perturbation
by the mean-flow correction, ${\rm i}\gamma\cU$, is then small
in the first equation in (\ref{x.3}), and if we again make the approximation
that $A$ is contant, we find
\be
\gamma \approx -A \int_0^\tau e^{-\lambda q^3/3 - {\rm i}q\eta} {\rm d}q,
\label{xxx.3}
\ee
which is plotted in figure \ref{diffo}.
At $\tau\ll1$, (\ref{xxx.3}) recovers the secular growth of
the linear non-dissipative critical layer (\emph{cf} (\ref{4.25})),
but over longer times,
this solution approaches a steady state, illustrating
how diffusion is able to saturate that growth before nonlinearity (and the
advective term  ${\rm i}\gamma\cU$) enters the fray.
Figure \ref{diffo} also illustrates how this dynamics does indeed
characterize the full modified model for larger values of the
diffusivity, demonstrating how the analytical solution
in (\ref{xxx.3}) agrees satisfyingly with numerical results
computed with $\lambda=5.3$. The steady state prediction from (\ref{xxx.3})
corresponds to the result of viscous critical-layer theory presented by
\cite{boulanger} for stratified tilted vortices (in which case, $\tau\rightarrow\infty$ in (\ref{xxx.3}) and the solution
  can be related to the Scorer function).

\begin{figure}
\begin{center}
\includegraphics[width=0.8\linewidth]{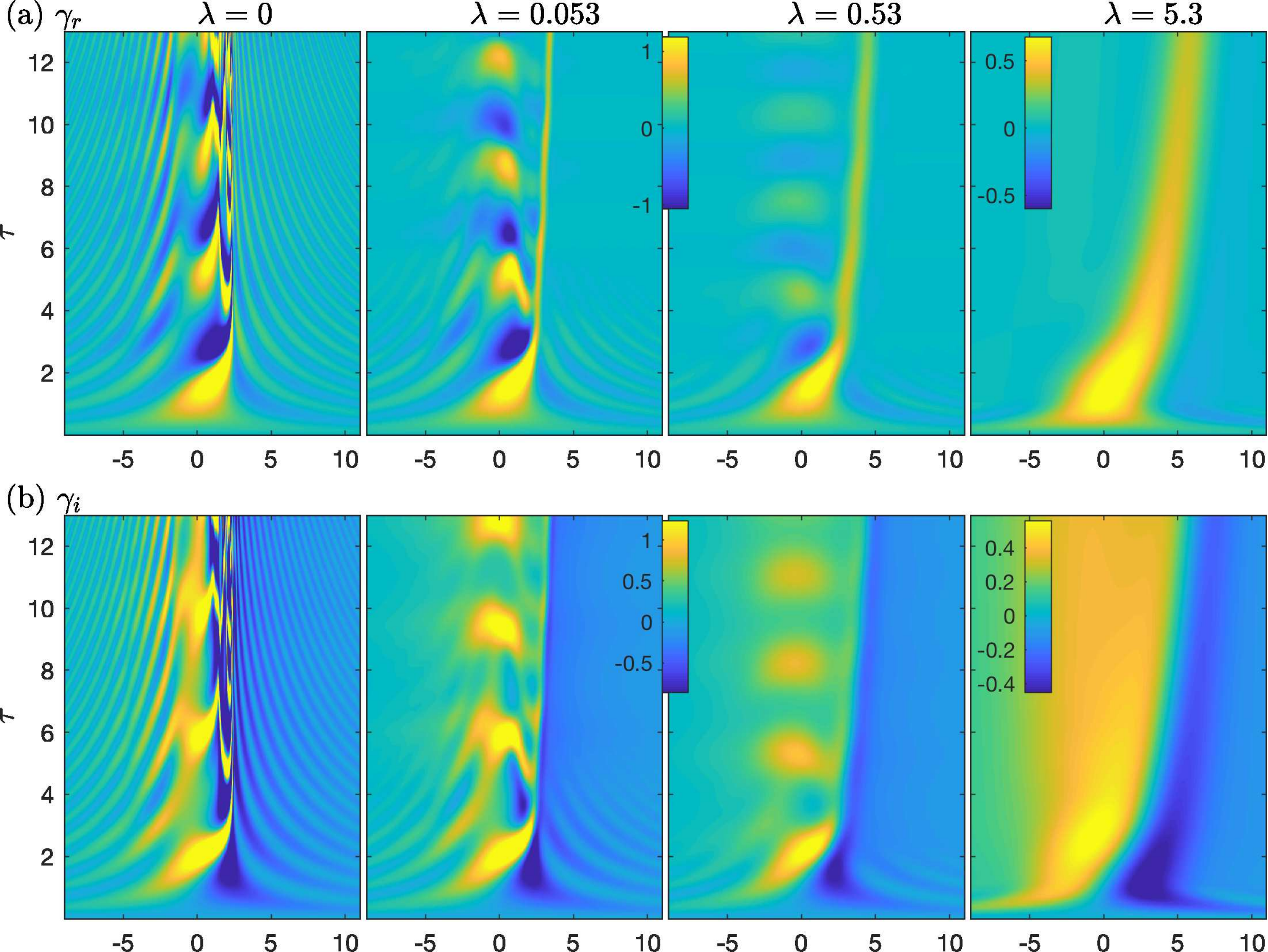}
\includegraphics[width=0.8\linewidth]{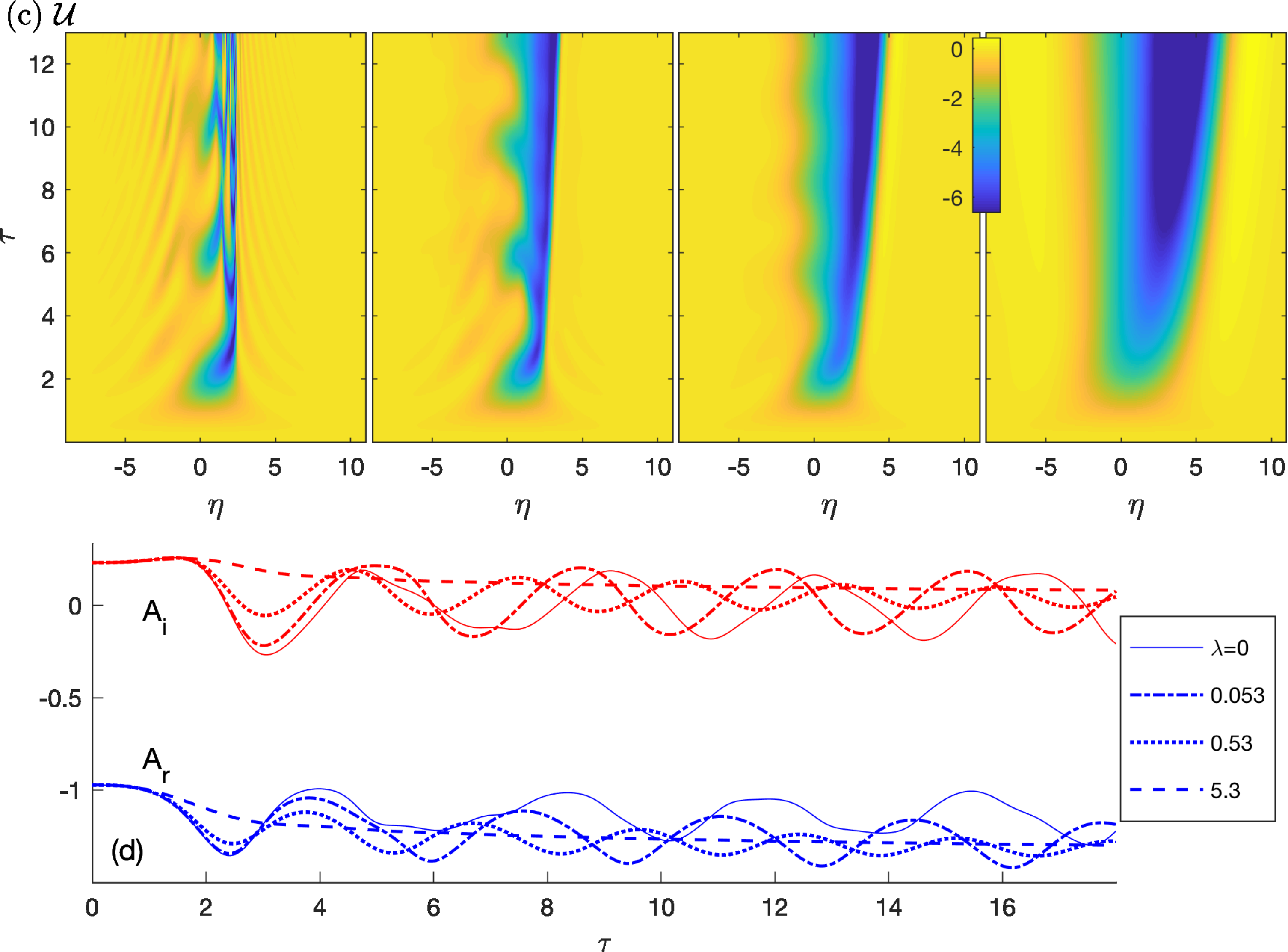}
\caption{ Solutions of the modified canonical model, showing
  (a) $\gamma_r$, (b) $\gamma_i$ and (c) $\cU$, for
  $m=1/2$, $f=4/3$ and $\cN=4/3$, $c_1=0.238$, $c_2=0.219$ with the values of
  $\lambda$ indicated (and corresponding to the three columns).
  The colormap is the same in the first three panels
  of (a) and (b), but not the rightmost panel.
  The quasi-steady wave amplitude $A=A_r+\mathrm{i}A_i$
  for the four computations is shown in (d).
}
   \label{sols}
 \end{center}
\end{figure}

Nevertheless, the establishment of a steady state with spatial
structure in the density perturbation is inconsistent with
the integral relation in (\ref{x.R}). Indeed, if $\gamma$ approaches a
steady state, $\cU$ continues to grow linearly with $\tau$, and
for times of order $\lambda^{1/3}$, the advective term
${\rm i}\gamma\cU$ can no longer be neglected in (\ref{x.3}),
heralding the onset of a different, more complicated phase
of evolution. Figure \ref{sols} shows a suite of numerical solutions,
illustrating this later evolutionary stage for cases with
stronger diffusion (right-hand panels),
and other examples with smaller $\lambda$ (left-hand panels).
For the latter, diffusion is too weak to arrest the linear
growth in the critical layer and nonlinear focussing begin to occur;
only when the spatial scale has reduced sufficiently
does the dissipative effect take hold to limit the exponential
amplification found for $\lambda=0$. At that stage,
a new phase of evolution again emerges, much like that found
for stronger diffusion. In particular,
the oscillations of the non-dissipative dynamics begin to
fade with time, and a localized coherent structure
emerges that drifts to larger $\eta$ under the
advective effect of the mean-flow correction. The structure
leaves in its wake an increasingly strong deficit in $\cU$,
which is permitted by the constraint in (\ref{x.R}) because
diffusion may continually
lower $\cU$ as long as the gradients of $\gamma$ remain finite.

\subsection{Dissipative coherent structures}

\def\dd{{\mathrm{d}}}

The drifting coherent structure
can be analyzed further owing to its
fine spatial scale and the relatively slow timescale over
which the system develops once the larger-scale transients
have subsided: assuming that $\lambda\ll1$ and $A$ is real and constant,
we search for a
  quasi-steady travelling wave solution in which
\be
\gamma \approx \gamma(\xi) \qquad {\rm and} \qquad
\xi =
%\frac{\eta-\eta_c }{\sqrt\lambda}
\frac{\eta}{\sqrt\lambda}
- \int c\; \dd\tau
%\textcolor[rgb]{0.00,0.59,0.00}{-c\tau}
,
  \label{x.5}
\ee
which characterizes a coherent structure with
  a length scale of $\sqrt{\lambda}\ll1$ and a drift velocity
  given by $c$. % $c\sqrt{\lambda}$}.
%where $c$ is the speed of the structure and $\xi=0$, or
%$\eta=\eta_*=\eta_c-\sqrt\lambda\int c \;\dd\tau$, denotes its
%position
%%(this object is spawned near the nonlinear
%%critical level, $\eta=\eta_c$).
Hence,
\be
-c\gamma' + {\rm i}\eta_*  \gamma + A \approx -{\rm i} \gamma \cU + \gamma''
\qquad {\rm and} \qquad
- c \cU' \approx
A^* \gamma + A \gamma^* \approx 2 A \gamma_r,
\label{x.6}
\ee
  where $\xi=0$, or
  $\eta_*=\sqrt\lambda\int c\;\dd\tau$, prescribes
  the center of the coherent structure.
  This fifth-order system may be solved subject to the far-field
  constraints that $\gamma$ and $\cU$ approach constant values
  as $|\xi|\to\infty$. In particular, since the coherent structure invades
  a region to the right in which $\gamma_r=\cU=0$, but
  $\cU$ remains finite to the left (see figure \ref{sols}),
  we demand the limits
\be
(\gamma_r,\gamma_i,\cU)\to\left\{\begin{array}{ll}
 (0,G_+,0) & {\rm for} \ \xi\to\infty,\cr
 (0,G_-,\Delta\cU)
  & {\rm for} \ \xi\to-\infty,
  \end{array}\right.
\label{x.7}
\ee
where $G_+=A\eta_*^{-1}$, $G_-=A(\eta_*+\Delta\cU)^{-1}$ and
$\Delta\cU$ is the jump in the mean flow across the structure. (\ref{x.7}) imposes six boundary conditions to (\ref{x.6}). One must also remove the translational invariance
of the system by imposing an additional constraint.
Thus, given $\eta_*$, we solve (\ref{x.6}) subject to those seven
conditions, treating $G_-$ and $c$ as unknown parameters (eigenvalues).
This furnishes localized
structures taking the form of ``pulses'' in $\gamma_r$
and ``fronts'' in $\gamma_i$ and $\cU$.
Note that,
as the coherent structure drifts to the right, $\eta_*$ increases,
corresponding to an evolution of the coherent structure, which is treated
parametrically in the quasi-steady approximation of
(\ref{x.5}) and (\ref{x.6}).

\begin{figure}
\begin{center}
\includegraphics[width=0.9\linewidth]{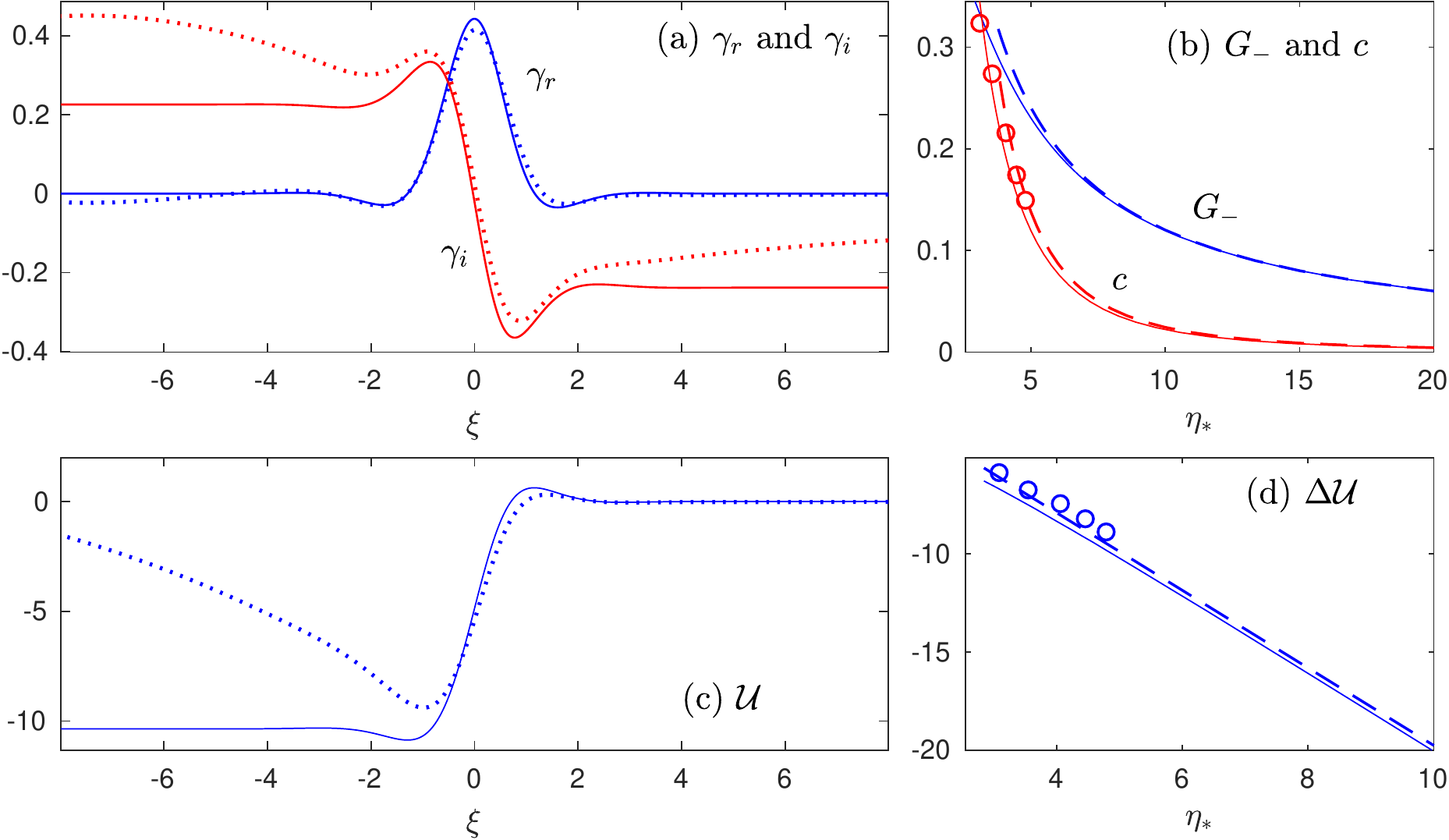}
\caption{
  A coherent structure computed from
  (\ref{x.6})
  with $\eta_*=5.04$ and $A=-1.2$, showing (a) $\gamma_r$ and $\gamma_i$,
  and then (c) $\cU$ (solid lines). The dotted lines show the numerical solution
  of the modified canonical model (\ref{x.3}),
  computed for $\lambda=0.53$ at $\tau=17.8$
  (at which moment the residual oscillations
  near $\eta=0$ are less pronounced).
  In (b) and (d) we show
  $G_-$, $c$ and the jump $\Delta\cU=-[\cU]_{-\infty}^\infty$
  against $\eta_*$ from the solutions to (\ref{x.6}) for $A=-1.2$ (solid lines).
  The dashed lines show the limiting behaviour for $\eta_*\gg1$ given in (\ref{x.8}). The circles show data for $c$ and $\Delta \mathcal{U}$ measured from the numerical solution of (\ref{x.3}) with $\lambda=0.53$ from $\tau=5$ to $\tau=15.4$.
}
   \label{soloplot}
 \end{center}
\end{figure}

Figure \ref{soloplot} shows a sample solution to (\ref{x.6})
for $(A,\eta_*)=(-1.2,5.04)$, giving $G_+=-0.24$.
These choices for $A$ and $\eta_*$ correspond to
the numerical solution of the modified canonical model for
$\lambda=0.53$ shown in figure \ref{sols} at $\tau\approx18$,
and they are also plotted in figure \ref{soloplot}.
The solution to  (\ref{x.6}) compares satisfyingly with the
snapshot of the simulations near the core of the coherent structure,
although there are discrepancies further away
  arising from the influence of the far-field flow.

Figure \ref{soloplot} also includes data computed from (\ref{x.6})
for $G_-$, $c$ and $\Delta\cU$, as functions of $\eta_*$. In the limit of large $\eta_*$,
a simple rescaling of (\ref{x.6}) and (\ref{x.7})) indicates
the limiting behaviour,
\begin{equation}
  G_-\to G_+ = O(\eta_*^{-1}),\quad c=O(\eta_*^{-5/2}),\quad \Delta\cU=O(\eta_*).
  \label{x.8}
\end{equation}
The solution of (\ref{x.6}) is compared to (\ref{x.8}) together with measurements from the numerical simulation
in the figure.
Similarly, the characteristic strength and width of the structure are $\gamma=O(\eta_*^{-1})$,
  $\mathcal{U}=O(\eta_*)$ and
$\xi=O(\eta_*^{-1/2})$.
Thus, as the coherent structure drifts to the right, and $\eta_*$
slowly increases, the drift velocity
  declines, and the
 peak in $\gamma_r$ and jump in $\gamma_i$
 must decrease and narrow.
 However, the jump in $\Delta\cU$ continues to
build up, predicting that the deficit in the mean flow
grows linearly with $\eta$ for $\eta<\eta_*$.

This behaviour of the coherent structure
rationalizes the dynamics of the modified canonical model
seen in figure \ref{sols}: once the linear dynamics and nonlinear focussing
have subsided, the two features that remain are the
decaying oscillations near $\eta=0$ and the
drifting coherent structure.
The structure leaves in its wake a slowly diffusing
density perturbation $\gamma\approx{\rm i}G_-$
(see the right-hand plots in figure \ref{sols}(b))
and a gradually strengthening mean flow correction $\Delta\cU$,
as seen on the right of figure \ref{sols}(c).
Thus, with diffusion, all growth in the density perturbation
becomes arrested, leaving a widening
and strengthening, jet-like defect in the mean flow.

One final concern is the impact of viscosity on the dynamics
of the coherent structure: it is clear from (\ref{x.2}) that
the growth of the mean flow correction may be halted when
$\nu=O(\varepsilon^2)$. Indeed, in the limit of stronger diffusion,
the viscous term may allow $\cU$ to also reach a steady state
within the critical layer. However, as for the classical critical
layers of Rossby waves \citep{Brown78} and clear from the constraint
(\ref{x.R}), a genuine steady state is not possible with dissipation.
Instead, the mean-flow correction must inevitably spread viscously
out of the critical layer, even if a quasi-steady
state is reached locally. Such considerations suggest that
viscosity, if sufficiently strong, may prevent the creation of the drifting
coherent structure, although a widening jet-like defect might still
appear in the mean flow.

\section{Discussion}

In this paper, we have studied the non-dissipative,
nonlinear dynamics of forced baroclinic
critical layers  using matched asymptotic expansion.
In the linear regime, the forcing establishes
a steady wave response outside the critical layers, but disturbances
grow secularly inside the critical layer, which thins with time.
The behavior is very similar to the forced critical layers of
both Rossby and internal gravity waves
\citep{Stewartson78,Warn76, Warn78,Booker67, Brown80}.
Continuing the analysis,
we then studied the weakly nonlinear dynamics of the critical layer,
finding that the adjustment of the mean flow provides the most
important feedback on the growing disturbance there. Guided by the
critical-layer scalings exposed by the weakly nonlinear analysis,
we then derived a reduced model for the nonlinear critical layer.
The numerical solution of the reduced model reveals
a continued growth of the vertical vorticity
as the disturbance is focussed exponentially quickly into
a finer region within the critical layer. The focussing progresses
uninterrupted until the reduced model breaks down.

Such pathological behaviour is quite different to that of the forced
critical layer of a Rossby wave, where nonlinearity halts the
secular linear growth and the mean vorticity distribution overturns
into a distinctive cat's eye structure
\citep{Stewartson78,Warn78,Mclntyre85}. In that process, all the
harmonics of the forcing pattern are excited to the same strength
of the fundamental component. By contrast, in our nonlinear
theory of the forced baroclinic critical layer,
the adjustment to the mean flow arrests the linear growth
and prompts the focussing of the vorticity
before any of the higher harmonics become important.
It is only once the strength and lengthscale of the focussed
vorticity pass out of the asymptotic regime of our theory that
the harmonics will appear.
One important contributor to this feature is that the position of
the baroclinic critical level itself is dictated the
streamwise wavenumber. The critical level of the
forcing does not therefore coincide with those of the harmonics.
 This filtering action
weakens the impact of those harmonics within the baroclinic critical
layer, leaving the adjustment the mean flow as the main nonlinearity.

The nonlinear structures developed in our forced baroclinic critical
layers (jet-like defects in the mean velocity
  and dipolar stripes in the vorticity)
 may well be the analogues
of features seen in the simulations
of Marcus \textit{et al.} (2013) and \citet{Wang16}.
Unlike in the reduced model, however, where
these structures continue to focus,
the mean flow structures spawned in the simulations
roll up into new vortices, providing part of the chain of events
leading to self replication. Thus, our model likely misses
important secondary instabilities. Indeed,
\citet{Mclntyre85} and \cite{Haynes89} have shown that the nonlinear evolution
of a forced Rossby wave can be susceptible to shorter-wavelength
shear instabilities and generate ``critical layer turbulence''
along the filaments of vorticity wrapped around the main cat's eye
(see also \cite{Balmforth01}).
A roll up of the jet-like defects into new vortices seems plausible
in the present case, and may arrest the uninterrupted focussing
effect within the nonlinear critical layer.
However,
an extension of the matched asymptotic analysis is required to
capture such dynamics.

  \citet{Marcus16} further argued that self replication is a finite-amplitude
  instability, requiring the amplitude of the initial disturbance to exceed
  a certain threshold. By contrast, the secular growth and nonlinear
  focussing of the disturbance inside the critical layer is triggered
  for an arbitrary small forcing amplitude in our analysis. Nevertheless,
  we have idealized the driving as a steady wavemaker, and ignored any
  possible evolution of that forcing. If the wavemaker cannot
  be sustained indefinitely, a threshold likely emerges that demands that
  the forcing act for sufficient time and strength to
  drive the baroclinic critical layers to the point where
  secondary instability can arise.

The continued focussing of the mean vorticity
layer also indicates that dissipative effects are likely to become
important in the later stages of evolution
inside the baroclinic critical layer. Including the diffusion of
density ({\it i.e.} heat or salt) in the theory
leads to a modification of the reduced model, which now takes
a partial differential form. A brief exploration of
the modified model demonstrates that weak diffusion can arrest
the focussing to the nonlinear critical level.
Interestingly, a drifting solitary-wave like object then
emerges, with a structure that can be analyzed analytically. The solitary
wave leaves in its wake another jet-like defect in the mean flow,
but this time the defect gradually widens and deepens as the object
drifts.

  In summary, when a steady forcing drives waves with baroclinic critical
  levels into a horizontally sheared flow with vertical stratification,
  the growing density perturbations predicted by linear theory
  become saturated by nonlinear effects. Although this saturation is
  demanded by the conservation laws of the governing equations,
  those constraints still permit the density perturbation to
  develop finer spatial structure over a region within the
  baroclinic critical layer. This nonlinear focussing effect
  takes place exponentially quicky, developing sharp jet-like defects in the
  mean flow, which can survive even in the presence of weak dissipation.
  This dynamics of the baroclinic critical layers is more destructuve
  than that for the classical critical layers of Rossby and internal
  waves, and plausibly rationalizes part of the cycle of vortex
  self replication observed by Marcus {\it et al.} in numerical
  simulations.

\acknowledgements
We thank Professors Philip Marcus, Richard Kerswell, St\'{e}phane Le Diz\`{e}s and Dr. Thomas Eaves
for important discussions, and the referees for helpful comments. We also thank Dr. Timm Treskatis and Mr. Mingfeng Qiu for help on our numerical simulation.
C.W. thanks the University of British Columbia for
 a Four-Year Doctoral Fellowship.

\appendix

\def\cV{{\cal V}}
\def\cU{{\cal U}}

\section{The critical-layer vorticity distribution} \label{appo}

The reconstruction of the critical-layer vorticity from the matched asymptotics
is:
\begin{equation}
  \zeta=\zeta_0+\left[\varepsilon^{\frac{1}{3}}\zeta_1e^{\mathrm{i}x+\mathrm{i}mz}
    +\varepsilon^{\frac{2}{3}}\zeta_2e^{2\mathrm{i}x+2\mathrm{i}mz}+\mathrm{c.c.}
    \right], \label{5.1}
\end{equation}
where $\zeta_0$ is given by (\ref{5.2}),
\begin{equation}
  \zeta_1=\mathrm{i}\left(\frac{m^2}{2\cN}\right)^{\frac{2}{3}}
  \left[(f-1)\gamma+\gamma\frac{\partial |\gamma|^2}{\partial \eta}
    + \frac{2}{m^2}{v_1}\frac{\partial^2 |\gamma|^2}{\partial \eta^2}\right],
    \label{5.3}
\end{equation}
\begin{equation}
  \zeta_2 = (f-1)\left(\frac{m^2}{2\cN^4}\right)^{\frac{1}{3}}
    \left(\frac{1}{2}m^2\gamma^2 - {{v_1}} \gamma_\eta\right)
    +\frac{\textrm{i}}{(2m\cN)^{\frac{2}{3}}}\frac{\partial}{\partial \eta}
    \left[v_1 \zeta_1+\left(\frac{m^4\cN}{4}\right)^{\frac{1}{3}}u_1\gamma\right],
    \label{5.4}
\end{equation}
and the leading-order fundamental components of the critical-layer
horizontal velocity are
\begin{equation}
  {v_1} = \frac{\mathrm{i}m^2A}{2}\left[\log\left|
    \left({\frac{\varepsilon^2m^2}{2\cN}}\right)^{\frac{1}{3}}\eta\right|+1\right]
  -\frac{m^2}{2}\fint^\infty_\eta \left(\gamma-\frac{\textrm{i}A}{\eta'}\right)
  \mathrm{d}\eta'+\frac{\mathrm{i} A(\alpha\cN-f)}{\cN^2-f(f-1)}, \label{A4}
\end{equation}
\begin{equation}
{u_1}= \frac{(f-1){v_1} -\mathrm{i}A+{v_1}\zeta_0}{\mathrm{i}\cN},
\end{equation}
where the decoration on the integral sign implies principal value.

\end{document}